\documentclass[trackchanges,twocolumn]{aastex701} 

\usepackage{amsmath}	
\usepackage{graphicx}	
\usepackage[table]{xcolor}
\usepackage{enumitem}
\usepackage{xspace}
\usepackage{multirow}

\definecolor{brightmaroon}{rgb}{0.7, 0.09, 0.09} 
\definecolor{lavendar}{rgb}{0.53, 0.50, 0.9} 

\definecolor{RLOF_color}{rgb}{0.9372549 , 0.67843137, 0.42352941} 
\definecolor{lowM_stripped_color}{rgb}{0.54509804, 0.80392157, 0.9372549} 
\definecolor{highM_stripped_color}{rgb}{0.501960784314, 0.592156862745,0.921568627451} 
\definecolor{transition_color}{rgb}{0.99215686, 0.45098039, 0.59215686} 
\definecolor{WD_color}{rgb}{0.88627451, 0.73333333, 0.78431373} 
\definecolor{CO_color}{rgb}{0.75686275, 0.89411765, 0.38431373} 

\newcommand{\Msun}{\ensuremath{\rm{M}_{\odot}}}

\newcommand{\MESA}{\texttt{MESA}\,}

\newcommand{\TotalSystems}{5,464}

\newcommand{\WUMabinaryCount}{684}

\newcommand{\highmassXRBCount}{162}
\newcommand{\HotsubdwarfbinaryCount}{146}
\newcommand{\AlgolCount}{126}
\newcommand{\PostAGBbinaryCount}{85}
\newcommand{\ChemicallyPeculiarCount}{82}
\newcommand{\ELCVnCount}{71}
\newcommand{\WRbinaryCount}{45}
\newcommand{\BluestragglerbinaryCount}{27}
\newcommand{\MassivecontactbinaryCount}{27}
\newcommand{\AstrometriccompactobjectCount}{25}
\newcommand{\SpectroscopiccompactobjectCount}{14}
\newcommand{\pulsarbinaryCount}{9}
\newcommand{\IntermediateMstrippedstarCount}{8}

\newcommand{\HegiantCount}{4}
\newcommand{\allWDMS}{3,949}

\begin{document}

\title{Ongoing and Post-Mass-Transfer Binaries:\\A Living Catalog and Unified Review of Binary Mass Transfer Products}

\author[orcid=0000-0001-5484-4987,sname='van Son']{Lieke A. C. van Son}
\affiliation{Department of Astrophysics/IMAPP, Radboud University Nijmegen, PO Box 9010, 6500 GL Nijmegen, The Netherlands}
\email[show]{lieke.vanson@ru.nl}  

\author[0000-0001-6970-1014]{Natsuko Yamaguchi}
\affiliation{Department of Astronomy, California Institute of Technology, 1200 E. California Blvd., Pasadena, CA, 91125, USA}
\email{nyamaguc@caltech.edu}

\author[orcid=0000-0002-1386-0603,gname=Pranav,sname=Nagarajan]{Pranav Nagarajan}
\affiliation{Department of Astronomy, California Institute of Technology, 1200 E. California Blvd., Pasadena, CA, 91125, USA}
\email{pnagaraj@caltech.edu} 

\author[0000-0003-0642-8107]{Tomer Shenar}
\affiliation{The School of Physics and Astronomy, Tel Aviv University, Tel Aviv 6997801, Israel}
\email{tshenar@tau.ac.il}

\author[0000-0002-8134-4854]{Koushik Sen}
\affiliation{Steward Observatory, Department of Astronomy, University of Arizona, 933 N. Cherry Ave., Tucson, AZ 85721, USA}
\email{ksen@arizona.edu}

\author[0000-0002-5522-0217]{Alexander Laroche}
\affiliation{David A. Dunlap Department of Astronomy \& Astrophysics, University of Toronto, 50 St George St, Toronto, ON M5S 3H4, Canada}
\affiliation{Dunlap Institute for Astronomy \& Astrophysics, University of Toronto, 50 St George Street, Toronto, ON M5S 3H4, Canada}
\email{alex.laroche@mail.utoronto.ca}

\author[0000-0002-3944-8406]{Emily M. Leiner}
\affiliation{Department of Physics, Illinois Institute of Technology, 3101 S. Dearborn St., Chicago, IL 60616}
\affiliation{Center for Interdisciplinary Exploration and Research in Astrophysics, Northwestern University, 1800 Sherman Ave, Evanston, IL, 60201, USA}
\email{eleiner@illinoistech.edu}

\author[orcid=0000-0001-6656-4130,sname='Sana']{Hugues Sana}
\affiliation{Institute of Astronomy, KU Leuven, Celestijnlaan 200D, 3001 Leuven, Belgium}
\affiliation{Leuven Gravity Institute, KU Leuven, Celestijnenlaan 200D, box 2415, 3001 Leuven, Belgium}
\email[]{hugues.sana@kuleuven.be}

\author[sname='Pols']{Onno R. Pols} 
\affiliation{Department of Astrophysics/IMAPP, Radboud University Nijmegen, PO Box 9010, 6500 GL Nijmegen, The Netherlands}
\email{fakeemail2@something.com}

\begin{abstract}
Mass transfer is arguably the most defining interaction in binary stellar systems, yet many aspects of its physics remain poorly understood, from stability to endpoints and observable products. 
Comparing theory and observations is challenging because post-mass-transfer systems are studied across largely independent communities with different methods, nomenclature, and evolutionary frameworks.
We present a unified review and catalog of ongoing and post-mass-transfer binaries spanning the full stellar mass range, but restricted to systems likely to have experienced only a single episode of mass transfer (i.e., only one component is evolved). 
We review 16 observational classes of binary interaction products and compile a curated sample of \TotalSystems\ systems into a publicly available, community-driven catalog at \url{https://binary-observations.github.io/post_mt_catalog/}.
Using this catalog, we investigate global trends in orbital periods, eccentricities, masses, and mass ratios across post-mass-transfer binaries. 
We find I) non-zero eccentricities are common at all periods and system classes, with both median values and scatter increasing with period, II) the $e(\log P)$ relation depends on donor progenitor mass, with neutron-star and black-hole binaries showing the highest median eccentricities, likely reflecting effects of natal kicks, III) period distributions are broad and overlapping across evolutionary channels, and IV) the Gaia BH and NS systems are extreme in mass ratio but otherwise consistent with the general post-mass-transfer population.
Together, these results support a unified empirical view of post-mass-transfer binaries that highlights several  tensions between theory and observations.
\end{abstract}

\section{Introduction \label{sec:introduction}} 
In the past decade, it has become increasingly clear that binary stellar systems are omnipresent \citep{2012Sci...337..444S,2023ASPC..534..275O,2024NewAR..9801694E}.
Binary mass transfer is arguably the most central and defining interaction in the life of a binary system.
It drives the formation of a wide variety of transients, ranging from Type~Ia supernovae \citep[e.g.,][]{2014ARA&A..52..107M}, and luminous red novae \citep[e.g.,][]{2017ApJ...835..282M}, to X-ray binaries \citep[e.g.,][]{2023A&A...671A.149F,2024A&A...684A.124F},  millisecond pulsars, and gravitational-wave sources \citep{2016Natur.534..512B, 2022ApJ...940..184V, 2026A&A...706A.296K}. 
On the low-mass end, binary interaction products such as blue stragglers, barium stars, and subdwarf~B stars have been known for decades, yet continue to pose fundamental challenges to theory \citep[see the recent review by][]{2025ARA&A..63..467M}.
On the high-mass end, interest has surged with the detection of gravitational waves from merging compact objects, systems whose formation hinges critically on one or more prior episodes of mass transfer \citep[see the review by][]{2024ARA&A..62...21M}.
Mass transfer is commonly classified as (dynamically) stable or unstable based on the donor star's response to mass loss and the orbital response to mass exchange.
If the donor can readjust while remaining within its Roche lobe, mass transfer proceeds stably, either on the donor's thermal or nuclear timescale \citep[e.g.,][]{1997A&A...327..620S}.
If instead the mass-loss rate runs away, the binary enters a common-envelope (CE) phase \citep[see reviews by][]{2013A&ARv..21...59I,2020cee..book.....I}, during which the companion spirals inward through the donor's envelope.
Despite significant recent progress with hydrodynamical simulations, CE evolution remains poorly understood, and most population synthesis models still rely on simplified energy-balance prescriptions \citep[e.g.,][]{1968IAUS...34..396P,1984ApJ...277..355W}.
Similarly, recent studies challenge key assumptions underlying stable mass transfer, including circular orbits, the conservativeness of mass transfer, and the accretor's response to mass gain \citep[e.g.,][]{2023ApJ...942L..32R,2026ApJ..1000....2S,2025A&A...695A.117N,2025ApJ...990L..51L,2025ApJ...983...39R,2026A&A...706A.357P}.

Significant discrepancies between theory and observations of mass-transfer outcomes persist and are not diminishing as modern surveys increase the sample of interacting and post-interaction binaries.
One example concerns the orbital-period distribution of white dwarf--main sequence (WD+MS) binaries: models predict either short-period systems ($P_\mathrm{orb} \lesssim 100$~d) from  common-envelope evolution, wider orbits from stable mass transfer, or very wide systems ($P_\mathrm{orb} > 10^4$~d) that avoided interactions altogether \citep[e.g.,][]{1984ApJ...277..355W,1976IAUS...73...75P,1997MNRAS.291..732T,1997A&A...327..620S,2017ApJS..230...15M}.\footnote{Note that wind interaction in even wider orbits is still possible \citep[as seen e.g. in the Mira AB system][]{2005ApJ...623L.137K}, but the effects on the evolution are much weaker.}
Over the past decade, expanding observational samples have shown that the predicted ``period gap'' between the post-CE and stable mass-transfer regimes is in fact populated across a range of system classes \citep{2018A&A...620A..85O,2019A&A...626A.127J,2024MNRAS.529.3729S,2024MNRAS.52711719Y,2024ApJ...970L..11H,2024A&A...686A..61B}.
More recently, the Gaia~BHs and NSs have posed a similar puzzle: main-sequence stars are in most cases orbiting the compact objects at separations that are too small to have avoided interaction, yet too wide to be readily explained by standard CE evolution \citep{2023MNRAS.518.1057E,2023MNRAS.521.4323E,2024A&A...686L...2G,2024OJAp....7E..58E}.
This has sparked a range of proposed solutions, ranging from alterations to binary mass transfer to more exotic formation scenarios (see \S~\ref{ss: Gaia_CO} and \ref{ss: are Gaia BHNS outliers} for more discussion).

Another persistent challenge is orbital eccentricity in post-mass-transfer systems.
The Roche-lobe formalism that underpins most mass-transfer models \cite{1971ARA&A...9..183P} assumes circular orbits and synchronized rotation, based on the expectation that tidal forces will rapidly circularize and synchronize the orbit once mass transfer begins \citep[e.g.,][]{1975A&A....41..329Z,1977A&A....57..383Z,1981A&A....99..126H}.
Yet a growing number of systems that have presumably undergone mass transfer retain significant eccentricities \citep[e.g.,][and section \S \ref{ss: P-e diagram}]{2009Natur.462.1032M,2011Natur.478..356G,2009A&A...505.1221V,2015A&A...579A..49V,2018A&A...620A..85O,2019A&A...623A.128H,2022A&A...659A...9P,2024MNRAS.529.3729S,2025A&A...701A...9M}.

Understanding these discrepancies and improving mass-transfer models requires a global and systematic understanding of the observational sample of post-mass transfer systems.
We would like to directly compare our theoretical models to observations, but in practice this remains challenging because the post-mass-transfer nature of many systems is unclear or actively debated.
Moreover, post-mass-transfer systems are studied across largely independent communities, each employing different detection methods, nomenclature, and evolutionary frameworks.
This makes it difficult to identify the general patterns in periods, eccentricities, mass ratios, and component masses.

The goal of this paper is to bridge this gap by providing a review and compiling a comprehensive catalog of binaries that have presumably undergone mass transfer, drawn from a wide range of literature sources and spanning the full mass spectrum from low-mass red-giant-branch donors to massive stars that produce neutron stars and black holes.
We focus on systems that have likely experienced a \textit{single episode} of mass transfer, in other words systems where only one of the two stars has likely transferred mass to its companion, while this companion has not yet evolved to fill its own Roche lobe. 
This helps us focus on the outcome of mass transfer itself, without the added complexity of later evolutionary phases.

We describe the construction of the catalog and the choices made during its compilation in Section~\ref{sec:catalog_construction}.
In Section~\ref{sec:both_stars_alive} we discuss systems in which both stellar components are still `alive', proceeding broadly from lower- to higher-mass progenitors of the donor.
Section~\ref{sec: binaries_with_wd} focuses on systems where one component is a white dwarf, and Section~\ref{sec: NSBHbinaries} considers binaries containing a neutron star or black hole.
In Section~\ref{sec:correlations} we use the compiled catalog to investigate global trends across post-mass-transfer binaries, examining the distributions of orbital periods, eccentricities, component masses, and mass ratios.
We summarize and conclude in Section~\ref{sec:conclusions}.

The full catalog can be found at Zenodo \citep[10.5281/zenodo.20441987][]{vanson_2026_postmt_zenodo}. An interactive version of the catalog, along with the code used to compile and analyze the data, is publicly available at \url{https://binary-observations.github.io/post_mt_catalog/}. We intend it to serve as a living community resource.

\section{Catalog construction} \label{sec:catalog_construction}
In this work, we focus on systems that have likely experienced only a single episode of mass transfer, i.e. where at most one component has evolved. By restricting to such binaries, the catalog aims to provide the most direct view of the mass-transfer process, without the added complexity and uncertainties introduced by subsequent evolutionary phases.

Throughout this work, we define the star that has (presumably) donated mass as the secondary. The star that has (potentially) accreted mass as the primary. 
%
In addition to recording each system’s position in the sky (RA and DEC), we aim to include systems for which at least basic orbital and mass information is available. 
Specifically, we focus on binary systems with a measured orbital period, an estimate of the eccentricity, and at least some constraint on the component masses. 
The latter could be given directly in terms of a literature value for $M_1$, $M_2$, their projected lower limits ($M_1 \sin^3 i$, $M_2 \sin^3 i$), or the binary mass function. 
Note that not all of this information is always available for every system.
Whenever a parameter is not directly measured but inferred or otherwise assumed, we flag this using the \texttt{quality\_flags} field.
This includes e.g., the flag \texttt{assumed\_e} for systems whose eccentricities are fixed to zero \textit{a priori} in the underlying modelling (this applies to all W~UMa binaries \S\ref{ss:WUMa}, massive contact binaries \S\ref{ss:contact}), and the majority of the Algol sample (\S\ref{ss: algols}). 
Similarly, e.g., \texttt{assumed\_M2} and \texttt{min\_M2} mark systems where the donor mass is an assumed value or a lower limit. 

We next assign types to the binary components, as well as an overall observational binary class to the system.
The component types are summarized in Table \ref{tab:component_types}. We distinguish between evolutionary and observational types for each component.
Evolutionary types represent a best guess for the inferred evolutionary states, such as main-sequence star (MS), white dwarf (WD), neutron star (NS), black hole (BH) etc.
Observational types record phenomenological information of the component. This can be the observed spectral type such as ``O5~V'', or a descriptive classification such as ``WR'' or ``Be star''.  
Because spectral classifications are reported in the literature with varying levels of specificity, and often involve numerous subclasses, the Observational Type field is kept flexible on purpose.

\begin{deluxetable}{ll}
\tablecaption{Component-level type classification. The listed observational types are illustrative examples; the format is free-form.  \label{tab:component_types}}
\tablewidth{0pt}
\tablehead{
\colhead{Evolutionary type (1,2) } & \colhead{Observational type (1,2)}
}
\startdata
MS                      & O5 V, O9.5 III, B1 Ve etc. \\
HG                      & A7 III \\
RGB                     & G8 IV \\
AGB                     & M-type giant \\
He-star/stripped star & sdB, sdO, WR \\
WD                      & DA WD, DB WD \\
NS                      & compact object \\
BH                      & compact object \\
\enddata
\end{deluxetable}

\begin{figure*}
\centering
\includegraphics[width=\textwidth]{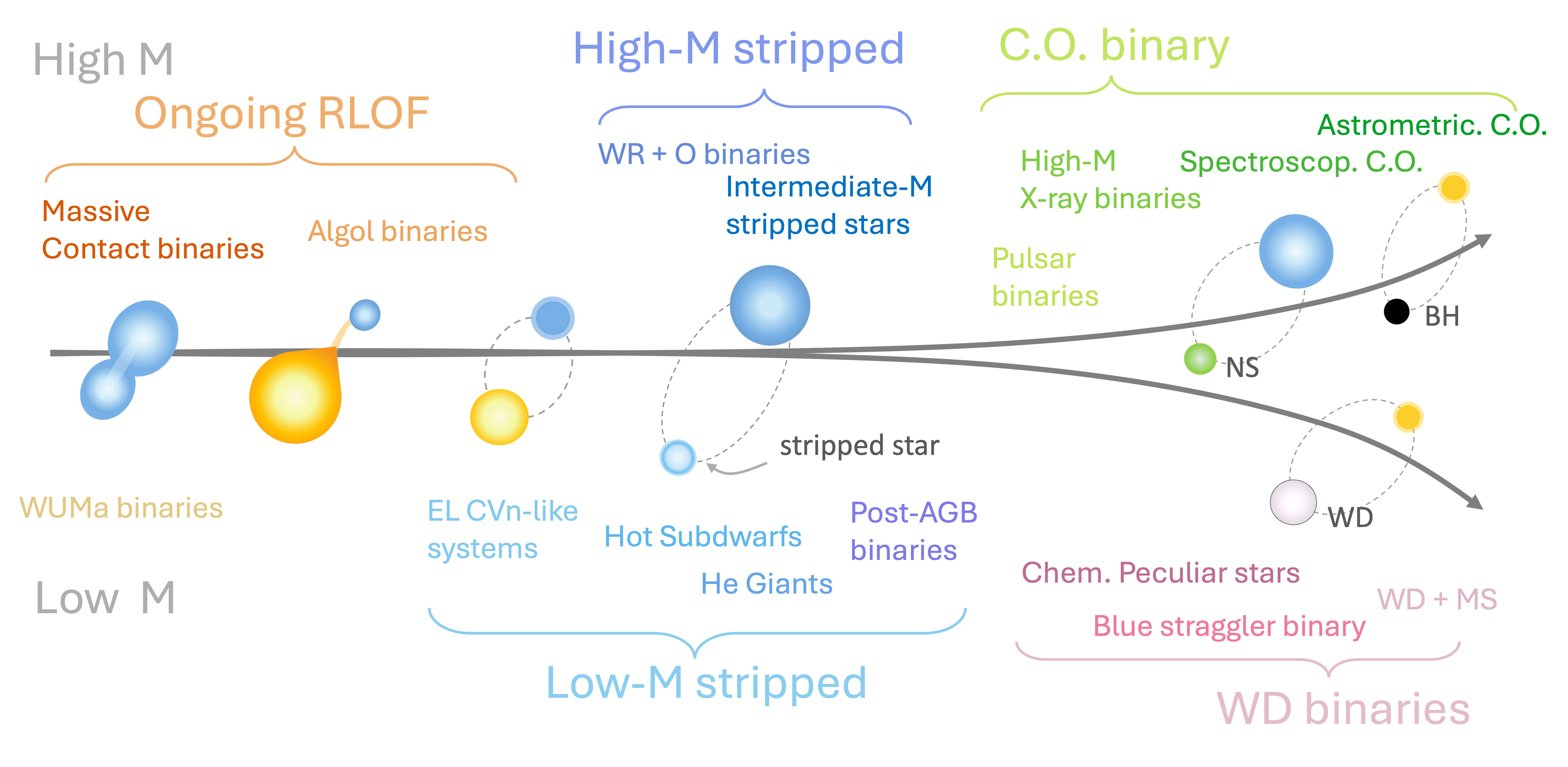} 
\caption{Cartoon depiction and overview of the post-mass-transfer binary system classes included in our catalog. Loosely organized by their evolutionary stage from left to right, and by mass (top vs bottom). We also indicate the category of mass transfer (RLOF, low-mass stripped, high-mass stripped, WD).}
\label{fig:overview_post_mt}
\end{figure*}

On the binary system level, we assign a `System class' to each binary system based on how it is identified or characterized in the literature.
Examples include Blue straggler stars,  Chemically Peculiar stars, Algol binaries, and X-ray binaries (see Table~\ref{tab:obs_classes}, and Figure~\ref{fig:overview_post_mt} ). 
These classifications link the system to commonly used system-level terminology under which these binaries are known, reported and discussed in the literature.
System classes also reflect the observational evidence for ongoing or past mass transfer, such as mass ratio reversal, stripped primaries, or chemical anomalies.
In some cases, the indicators of mass transfer are intrinsically linked to the system class itself (e.g., mass ratio reversal in Algol binaries), while in other cases multiple mass-transfer indicators may be relevant.

We include the literature references used for the catalog entry (given as an ADS bibcode\footnote{Each bibcode directly links to the SAO/NASA Astrophysics Data System via \url{https://ui.adsabs.harvard.edu/abs/<ADS bibcode>/abstract}. 
The full list of references is also available at \url{https://ui.adsabs.harvard.edu/public-libraries/Zm5P-O4cQPChveuI79eCzA}.}), a link to the \texttt{SIMBAD} database generated from the recorded RA and Dec, and free-form notes when relevant.

\paragraph{Duplicate handling}
We identify duplicate entries based on their RA and DEC. Systems separated by less than 0.1 arcsecond are considered identical, and their entries are merged. In these cases, we retain the most complete entry: i.e., the one where most fields are filled in, and combine the references from both entries. Additionally, for all systems we cross-check the recorded positions in \texttt{SIMBAD} and include a link to the corresponding \texttt{SIMBAD} page for each entry.
We further enrich our catalog by searching for spectral types in \texttt{SIMBAD} for all entries where this information is missing. We include the retrieved spectral types in the Observational Type field for the respective component.
Lastly, we include a \texttt{Notes} field for each entry, where we record any additional information that may be relevant for the interpretation of the system.

\subsection{Online catalog tools }
Our catalog is publicly available at \url{https://binary-observations.github.io/post_mt_catalog/}, and we intend it to be a living resource that the community can contribute to in the future.
Users can interact with the catalog through a web interface, which allows for easy browsing and searching of the entries.
The catalog is also available for download in various formats (e.g., CSV, JSON) for further analysis and use in other applications \citep[10.5281/zenodo.20441987][]{vanson_2026_postmt_zenodo}.
We welcome contributions from the community; users can suggest missing systems and propose new system classes through the GitHub repository.\footnote{\url{https://binary-observations.github.io/post_mt_catalog/missing_systems.html}}.

\begin{table*}[htbp]
    \centering

    \hspace{-2cm}
    \begin{tabular}{lllll}
        \hline \hline
        \textbf{Category} & \textbf{System class}  & \textbf{Description}  & \textbf{Components} & \textbf{N} \\
        \hline \hline
        \multicolumn{4}{c}{\textit{Binaries with two alive components}} \\
        \hline
        \multirow{2}{*}{\rotatebox{90}{\parbox{1cm}{\centering \color{RLOF_color} RLOF}}} 
        & \ref{ss:WUMa} WUMa binaries        &
            Both stars fill Roche lobe &
            MS + MS &  \WUMabinaryCount \\
        & \ref{ss:contact} Massive Contact binaries        &
            Both stars fill Roche lobe &
            MS + MS &  \MassivecontactbinaryCount \\
        & \ref{ss: algols} Algol binaries                        &
            Semi-detached, mass ratio reversal &
            MS + MS & \AlgolCount \\
        \multicolumn{4}{c}{\dotfill} \\
        \multirow{3}{*}{\rotatebox{90}{\parbox{1.3cm}{\centering \color{lowM_stripped_color} low M stripped}}} 
        & \ref{ss: EL CVn}  EL CVn-like systems                    &
            Pre--He WD phase with MS companion &
            pre-He-WD + MS & \ELCVnCount \\
        & \ref{ss: low M ss} Hot Subdwarfs              &
            core or shell He-burning stripped stars&
                sdB/sdO + MS & \HotsubdwarfbinaryCount\\
        & \ref{ss: low M ss} He giants                  &
            likely case AB mass transfer  &
                He giant + MS & \HegiantCount\\
        & \ref{ss: post AGB} Post-AGB binaries                     &
                Recently terminated MT  &
                core of AGB + MS &  \PostAGBbinaryCount \\

                \multicolumn{4}{c}{\dotfill} \\

        \multirow{3}{*}{\rotatebox{90}{\parbox{1.cm}{\centering \color{highM_stripped_color} high M stripped}}} 
        & \ref{ss: intermediat M stars} Intermediate-M stripped stars      &
            stripped star mass around $1.5$–$8\Msun$ &
                stripped star + MS &  \IntermediateMstrippedstarCount \\
        & \ref{ss: WR binaries} WR + O binaries                       &
            Wolf-Rayet + massive companion &
            WR + MS & \WRbinaryCount \\

                \multicolumn{4}{c}{\hspace{\linewidth}} \\
        \hline
        \multicolumn{4}{c}{\textit{Binaries containing a white dwarf}} \\
        \hline
        \multirow{3}{*}{\rotatebox{90}{\parbox{1.5cm}{\centering \color{WD_color} WD}}} 
        & \ref{ss: blue_stragglers} Blue straggler stars                  &
            rejuvenated companions &
            WD + BSS &  \BluestragglerbinaryCount \\
        & \ref{ss: chemically_peculiar_stars} Chemically peculiar stars             &
            Barium, CEMP, lithium-enhanced stars &
            WD + chem. pec.  & \ChemicallyPeculiarCount \\
        & \ref{ss: WD + MS} WD + MS ($P \lesssim 10^{3}\mathrm{d}$)    &
            Spectroscopic, Astrometric, Self-lensing &
            WD + MS &  \allWDMS \\
        \hline
        \multicolumn{5}{c}{\textit{Binaries containing a neutron star or black hole}} \\
        \hline
        \multirow{4}{*}{\rotatebox{90}{\parbox{1.8cm}{\centering \color{CO_color} C.O. binary}}} 
        & \ref{ss: pulsar_binaries} Pulsar binaries                 &
            With non-degenerate companion &
            NS + MS  &  \pulsarbinaryCount \\
        & \ref{ss: XRB} X-ray binaries (XRBs)                 &
            high-mass XRBs &
            NS/BH + MS  & \highmassXRBCount \\
        & \ref{ss: spectroscopic_CO} Spectroscopic compact object &
            Detached binaries; NS/BH companion &
            compact object & \SpectroscopiccompactobjectCount \\
        & \ref{ss: Gaia_CO} Astrometric compact object        &
                Gaia binaries, NS/BH companions &
                compact object & \AstrometriccompactobjectCount \\

        \hline
    \end{tabular}
    \caption{
        Observational binary system classes and their typical components. Over-arching categories are indicated in colored text on the left. The number of systems in each class included in our catalog is shown in the rightmost column. The full catalog is available at \url{https://binary-observations.github.io/post_mt_catalog/}.
        \label{tab:obs_classes}
    }
\end{table*}




\section{Binaries with alive components}
\label{sec:both_stars_alive}

In the following subsections, we describe the system classes in our catalog in which both binary components are ``alive''. In general, this means that both stars are still undergoing nuclear burning and/or are non-degenerate. I.e, the `Ongoing RLOF', `low-mass stripped' and `high-mass stripped' categories in Figure~\ref{fig:overview_post_mt} and Table~\ref{tab:obs_classes}. We nevertheless include the edge cases of post-AGB binaries (\S~\ref{ss: post AGB}) and EL CVn-like systems (\S~\ref{ss: EL CVn}), in which the donor is transitioning to a WD. We classify these systems as binaries with alive components because they are so recently post-mass-transfer that the donor has not yet fully contracted into a WD, but we note that this is a somewhat arbitrary choice.

\subsection{W~UMa contact binaries \label{ss:WUMa}}
Contact binaries are systems in which both stars fill or overfill their Roche lobes, sharing a common envelope and undergoing ongoing exchange of mass and angular momentum.
At low masses, contact binaries manifest as the W~Ursae~Majoris (W~UMa) class \citep[for reviews see e.g.,][]{2003ASPC..293...76W}. 
W~UMa systems are characterised by short orbital periods (${\sim}0.2$--$1$\,d), light curves with eclipses of nearly equal depth, and components of unequal masses ($q\sim 0.3$) but nearly identical surface temperatures ($\Delta T \lesssim 200$\,K). The latter is caused by the energy redistribution across the shared envelope \citep{1968ApJ...151.1123L,2005ApJ...629.1055Y}.
W~UMa progenitors are thought to reach contact through angular-momentum loss acting on initially detached, tidally-locked low-mass close binaries, via magnetic braking from magnetised winds \citep[e.g.,][]{2006AcA....56..199S,2011AcA....61..139S} and, as suggested by the high incidence of tertiary companions, Kozai--Lidov cycles and tertiary-driven angular-momentum loss \citep[e.g.,][]{2006AJ....131.2986P}.
The subsequent contact-phase evolution is expected to be governed by a combination of further loss of angular momentum, internal mass transfer, and energy transfer through the common envelope 
\citep{1976ApJ...205..217F,2005ApJ...629.1055Y}.
Continued angular-momentum loss drives the orbit toward smaller separation and the mass ratio toward more extreme values; once q falls below a critical threshold $q_{\rm min} \sim 0.07$ (see however \citealp{2024A&A...692L...4L}), the Darwin instability ensues, terminating the contact phase in a merger event \citep{1995ApJ...444L..41R, 2001ApJ...562.1012E,2009MNRAS.394..501A}. 
The merger is accompanied by L2 mass loss and most likely a luminous red-nova outburst \citep[e.g., V1309 Sco ][]{2011A&A...528A.114T,2016A&A...592A.134T}. The merger remnant is likely a blue straggler, consistent with the over-representation of W UMa-like progenitors in the blue-straggler populations of old open clusters \citep[e.g.,][]{1990AJ....100..469M,2009Natur.462.1032M}.

We note that orbits of W~UMa systems are circular by construction: the Roche-lobe geometry on which the underlying light-curve solutions are built assumes synchronous rotation and $e=0$. 
We include the $684$ W~UMa systems compiled by \cite{2021ApJS..254...10L}, available through the online catalog \texttt{WUMaCat}\footnote{\url{https://wumacat.aob.rs}}.
This is the largest published compilation of W~UMa stars with full light-curve solutions, and supersedes the earlier compilations \citep[e.g.,][]{2003CoSka..33...38P, 2004A&A...426.1001C,2011MNRAS.412.1787D}.

\subsection{Massive contact binaries \label{ss:contact}}
In massive contact binaries, the shared envelope is largely radiative, and most known systems are found close to the zero-age main sequence.
\cite{2024A&A...691A.150V} measured the rate of period change for a sample of B+B and O+B overcontact systems, finding timescales consistent with nuclear-timescale evolution.
Population synthesis studies indicate that about $1\%$ of all O-type stars are expected to be found in a contact configuration \citep[e.g.,][]{2021MNRAS.507.5013M,2024A&A...682A.169H}.
The evolutionary outcome of massive contact binaries is thought to be governed by mass loss through the outer Lagrangian point: once the envelope overflows $L_2$, the system is expected to enter a common-envelope phase, often leading to a stellar merger \citep{2021MNRAS.507.5013M}, although in some cases the system may survive as a close binary \citep[e.g.,][]{2016A&A...588A..50M}.
Efficient rotational mixing in massive, tidally locked systems can drive chemically homogeneous evolution \citep[CHE; e.g.,][]{2009A&A...497..243D,2013A&A...556A.100S,2020A&A...641A..86H}.

Massive contact binaries are typically identified by their short orbital periods (${\sim}0.5$--$3$\,days) and large-amplitude, smooth, sinusoidal light curves \citep{2025CoSka..55c.390A}.
However, distinguishing true contact systems from near-contact or semi-detached configurations can be difficult \citep{2026A&A...705A..43M}.
The most definitive way to confirm that a system is in contact is to directly resolve the stellar surface (e.g., via interferometry), but this is not feasible for the vast majority of known systems.
The next most reliable method is a simultaneous fit to both the light curve and radial velocity curve using an eclipsing binary modelling code such as \textsc{phoebe} \citep{2005ApJ...628..426P,2020ApJS..250...34C} and/or \textsc{spamms} \citep{2020A&A...636A..59A}.
\cite{2025CoSka..55c.390A} note that, to date, only 13 massive contact systems have been conclusively confirmed via a combined light-curve and radial-velocity fit (see their Table~1).

We include all systems from \cite{2025A&A...695A.223V,2025A&A...695A.109F}, which provide the observed massive (near-)contact systems in the Milky Way, M31, and the Magellanic Clouds (in their Table~B1 and Tables~D1 and D2, respectively). This encompasses the 13 systems from \cite{2025CoSka..55c.390A}.
We note that both \cite{2025A&A...695A.223V} and \cite{2025A&A...695A.109F} \textit{assume} circular orbits ($e = 0$) for all contact systems, following the expectation that tidal forces circularise close binaries on short timescales \citep{1975A&A....41..329Z}.
We adopt their assumption and set $e = 0$ for all contact binaries, and indicate this with the \texttt{assumed\_e} quality flag, but caution that the listed eccentricities should be understood as modelling assumptions rather than measurements.

\subsection{Algol binaries \label{ss: algols}}
Algol binaries are semidetached systems in which the less massive star is more evolved than its companion \citep[the classic ``Algol paradox'', see ][and references therein]{1998A&AT...15..357P}. They are short-period and eclipsing binaries, enabling accurate dynamical mass measurements of the stellar components \citep{2004A&A...417..263B,2004yCat.5115....0S}. The less massive star fills its Roche lobe, providing clear evidence for ongoing mass transfer. The mass-gaining star has the luminosity and radius of a main-sequence star at its current mass, while the donor is overluminous for its mass \citep[e.g.,][]{2022A&A...659A..98S}. 

Algols are canonical examples of stable Roche-lobe overflow on the main sequence, as observational counterparts to the slow Case\,A mass transfer \citep[e.g.,][]{2001ApJ...552..664N,2007A&A...467.1181D,2017A&A...599A..84M}. 
Observationally, Algols are most commonly identified as eclipsing and double-lined spectroscopic binaries, with photometric variability as the primary discovery channel and radial velocity curves enabling detailed characterization. Contact binaries share these signatures but are distinguished from Algols by their light-curve shape.
Their eclipses and double-lined radial velocities allow for precise mass and radius determinations.
We note that a related class of interacting Algol systems, the so-called `Double Period Variables', exhibit an additional long photometric cycle on timescales of $\sim30-35$ times the orbital period \citep{2003A&A...399L..47M}. This superorbital variability is often attributed to cyclic changes in an accretion disc surrounding the accretor, potentially driven by variations in the mass-transfer rate, although its physical origin remains uncertain \citep[e.g.,][]{2025A&A...701A..90G}. We do not include these systems in the present work.

We include the catalog of 119 semi-detached, double-lined eclipsing binaries with well-determined absolute parameters from \cite{2020MNRAS.491.5489M}, which was compiled from \cite{2006MNRAS.373..435I}, the Catalogue of Algol-Type Binary Stars \cite{2004A&A...417..263B}, and \cite{2004yCat.5115....0S}. 
Additionally, we include all the massive Algol systems listed in \cite{2026ApJ..1000....2S}. 
As with the contact binaries (\S\ref{ss:WUMa},\S\ref{ss:contact}), we set $e=0$ unless otherwise specified, because the Roche-lobe-geometry light-curve solutions typically used, assume circular and synchronous rotation by construction. 
The eccentricities catalogued for these systems are thus \textit{not} an independent measurement, and we indicate this with the \texttt{assumed\_e} quality flag.
Nevertheless, substantial eccentricity would show up as unequal spacing between minima in the lightcurve, producing a poor fit to any circular-orbit model. We thus expect the sample to be largely circularized.

\subsection{Low-mass stripped stars \label{ss: low M ss}}
We consider systems to be a low-mass stripped star if the  presumed stripped star (presumed past donor) is currently burning helium in its core or shell, and has a mass below roughly $1.5\Msun$. 
This definition corresponds roughly to low- and intermediate-mass progenitors (i.e., $M_{\rm{ZAMS}} \lesssim 8\Msun$).

\subsection{EL CVn systems \label{ss: EL CVn}}
EL~CVn systems are short-period post–mass transfer binaries consisting of a bloated, hot, low-mass stripped star in the pre–helium white dwarf phase and an early-type (A/F) main-sequence companion. 
They have orbital periods of roughly $0.5-4$ days and are mostly identified as eclipsing and/or spectroscopic binaries through combined photometric and radial-velocity observations, from which their absolute system parameters are determined. 
These systems are typically interpreted as the remnants of stable mass transfer \citep[e.g.,][]{2017MNRAS.467.1874C}.

We include the 29 EL~CVn binaries from \citet{2025ApJ...979..108X} for which orbital and stellar parameters were derived that were first identified in the \textit{TESS} eclipsing binary catalog, as well as 36 eclipsing EL~CVn systems listed in \citep{2018MNRAS.475.2560V}, discovered by the Palomar Transient Factory.
We further include the six binaries characterized by \citet{2026A&A...705A.146S}, which were originally proposed as BH candidates based on their Gaia DR3 spectroscopic orbits \citep{2022MNRAS.515.1266E}. 
Spectral disentangling reveals cool ($\sim5\,000$~K), bloated, very low-mass ($0.19$--$0.28\,\Msun$) stripped stars with rapidly rotating A/F-type accretors ($1.9$--$2.2\,\Msun$) in $10$--$15$ day orbits. These donors are too low in mass to ignite He and have only recently detached: they are proto--He white dwarfs that will contract and heat through the (hot) pre--He~WD phase probed by the EL~CVn systems, and we therefore list them under this class.




\subsubsection{Hot subdwarfs (with alive companions) \label{ss: SdOB} }
Hot subdwarfs (sdOB) are hot, low-mass stars at the blue end of the horizontal branch \citep[see e.g., the recent review by][and references therein]{2026enap....2..488H}.
They are generally interpreted as the stripped remnants of red-giant stars that have lost their envelope in binary interactions. 
SdB stars are typically core helium burning, while more evolved sdO stars may exhibit He shell burning. 
%

Most of the hot subdwarfs included in this work are drawn from the catalog of \citet{2021ApJ...920...86K}, which was compiled to identify post–common-envelope binaries. They require (1) the presence of a hydrogen-depleted component that has lost its envelope, and (2) published observational constraints on fundamental system parameters, including orbital periods with $P \leq 100\,\mathrm{d}$, and mass estimates. While their period cut is intended to exclude systems that experienced no-, or stable mass transfer, we expect that some systems in this catalog may still be post–stable mass-transfer binaries. 
Since we primarily focus on systems post a \textit{single} mass-transfer phase, we exclude all systems consisting of a sdOB + WD/NS/BH from this sample. 
We additionally make sure all 23 systems from \cite{2019MNRAS.482.4592V} with known orbital periods, masses and eccentricities are included, as well as  all sdOB systems with Be-star companions listed in \citet{2025ApJ...990L..51L}. The presence of a Be companion provides strong evidence for past mass transfer, as the rapid rotation of Be stars has often been suggested to come from spin up through accretion \citep[e.g.,][]{1991A&A...241..419P,2020A&A...641A..42B,2026enap....2..430R}

\subsubsection{Helium giants \label{ss: He giants}}
There is a small class of `helium giants', (also referred to as hydrogen-deficient binaries), which are thought to be the stripped remnants of intermediate-mass binary systems. 
Their formation scenario involves an initial phase of mass transfer on the red-giant branch, followed by a second phase that removes the remaining hydrogen envelope, thereby exposing the CNO-processed helium core \citep[e.g.,][]{2006epbm.book.....E}. 
Only four of such systems are currently known \citep{2023MNRAS.518L..75J}: Upsilon Sagittarii \citep{2023MNRAS.518.3541G,2023MNRAS.518L..75J}, KS Per \citep{1982A&A...113L..22D}, LSS 4300 \citep{1984ApJ...276..229S}, and LSS 1922 \citep{1980ApJ...242L..43D}.

\subsection{Post-AGB binaries \label{ss: post AGB}}  
Post-asymptotic giant branch (post-AGB) binaries are systems in which the donor star has recently evolved off the AGB (or, for a substantial fraction, the RGB) and is on its way to becoming a white dwarf \citep[see e.g., review by][]{2003ARA&A..41..391V}.
Observationally, they are often identified by infrared excesses in their spectral energy distributions, which are attributed to the presence of circumbinary dust surrounding the binary.  
The short duration of the post-AGB phase ($\sim10^{3}$–$10^{5}$ yr), as well as the short dissipation timescale of the circumbinary disk \citep[about $10^4$ years e.g.,][]{2023MNRAS.521...35I}, implies that the binary interaction must have occurred recently. 
%

We include the 85 Galactic post-AGB systems from the catalog compiled by \cite{2022A&A...658A..36K}, and cross match them with the binaries with known orbital solutions from \cite{2018A&A...620A..85O}. Finally we supplement the orbital data from table C1 from \cite{2025A&A...703A.294M}. 

\subsection{Intermediate-mass stripped stars \label{ss: intermediat M stars}}

Intermediate-mass stripped stars are those with current masses around approximately $1.5$–$8\Msun$, placing them between the lower-mass sdOB stars and the higher-mass Wolf–Rayet stars. They are expected to be the descendants of roughly $8$–$20\Msun$ donor stars.\footnote{Though note that this range is highly metallicity dependent. } 
While models predict such objects to be common, they have long remained elusive observationally. 
A recent breakthrough was made by \citet{2023Sci...382.1287D,2023ApJ...959..125G}, who identified a sample of these systems through excess UV emission and follow-up spectroscopy, beginning to fill this gap. 
Prior to these discoveries, the only well-established example was HD~45166 \citep{2008A&A...485..245G}, but recent work suggests that this system is the merger product of two lower-mass helium stars \cite{2023Sci...381..761S}, and therefore may not be a post-mass-transfer binary.
Although all systems with multiple-epoch observations in the samples of \citet{2023Sci...382.1287D,2023ApJ...959..125G} show radial-velocity variations, indicating the presence of a companion, full orbital solutions are not yet available. 
As a result, we do not include these systems in our catalog at this stage.

Instead, we include a set of systems that were identified through targeted studies of individual objects rather than large-scale survey searches. 
Firstly, there have been several systems that were initially classified as containing a BH, but were later reinterpreted as binaries containing stripped stars, including LB-1 \citep{2020A&A...639L...6S}, and NGC 1850 BH1 \citep{2022MNRAS.511L..24E}. 
Two more such systems host bloated stripped stars that have only very recently ceased mass transfer and are still undergoing short-lived structural readjustment phases: VFTS 291 \citep{2023MNRAS.525.5121V} and $\gamma$ Columbae \citep{2022NatAs...6.1414I}. 
We further include SMCSGS-FS 69 \citep{2023A&A...674L..12R}, a shell-hydrogen-burning stripped star in the Small Magellanic Cloud with a Be-star companion, as well as three partially stripped star + Be/Oe binaries in the Magellanic Clouds—2dFS 163, 2dFS 2553, and Sk$-$71 35—identified by \citet{2024A&A...692A..90R}, with stripped-star masses in the range $4$–$8\Msun$.

\subsection{Wolf-Rayet + OB star binaries \label{ss: WR binaries}}
WR stars are an observational class of hot stars exhibiting broad emission lines in their spectra due to powerful stellar winds \citep[see e.g.,][and references therein]{2026enap....2..569S}. 
Classical WR stars are massive and hydrogen-depleted, indicating that they have lost their envelope.
It has long been debated whether this envelope stripping is caused by strong winds in single-star evolution \citep{2007ARA&A..45..177C}, or if WR stars are products of binary interaction \citep{1967AcA....17..355P, 1998NewA....3..443V}.
Potentially both channels contribute, with their relative importance depending on metallicity \citep{2020A&A...634A..79S},
though multiple studies suggest that also apparent single WR stars may be binary products \citep{2015A&A...581A..21H,2016A&A...591A..22S,2018A&A...611A..75S, 2024A&A...689A.157S}. 
Recently, \cite{2025A&A...695A.117N} used the observed masses and orbital periods of 21 WR+O binaries to constrain the mass-transfer efficiency and  angular-momentum loss, finding that most systems are best explained by highly  non-conservative Case~A mass transfer. They also noted that post-Case~B systems  appear significantly underrepresented in the observed WR+O population, suggesting either an intrinsic preference for Case~A origins or severe  observational selection effects against wider post-Case~B systems.

 In this work we focus on WR+OB binaries, which are strong candidates for past interactions. 
 WR+O binaries are by far the most common type of WR binary known. 
 There are only a few known WR+B binaries \citep{2001NewAR..45..135V} and one established case of a WR+BH binary -- the X-ray binary \object{Cyg~X-3} \citep[][see also \S \ref{ss: XRB}]{2017MNRAS.472.2181K}. 
 The majority of WR+O systems span periods of days to months, though periods up to few years are sometimes seen \citep[e.g.,][]{2003MNRAS.338..360F, 2016MNRAS.461.4115R, 2016A&A...591A..22S, 2019A&A...627A.151S, 2019MNRAS.488.1282W, 2022A&A...664A..93D}.

 We retrieved all WR+OB binaries with an established period in the MW, LMC, and SMC, most of which are listed in \citet{2001NewAR..45..135V, 2019A&A...627A.151S, 2016A&A...591A..22S}, respectively. We omit short-period WR binaries ($P \lesssim 1\,$d) without established spectral types for companions  due to possible confusion with intrinsic variability of the WR star. We used updated literature \cite[e.g.][]{2020A&A...641A..26D, 2022A&A...664A..93D, 2023A&A...674A..88D}, when applicable. We do not include WR stars suspected to be very massive main-sequence stars (e.g., WR~22). We also omit binaries in which only a photometric period was reported without a spectroscopic confirmation (e.g., WR~50), for the same reason.

\section{Binaries containing a white dwarf}
\label{sec: binaries_with_wd}
%
We discuss several classes in which the donor star is most likely to have become a WD. 
Though the presence of a WD companion is often supported by indirect evidence (e.g., UV excess or evolutionary signatures), in many cases the presence of a WD remains a working hypothesis rather than a confirmed observational result.
The system classes we include in this category are blue stragglers (\S\ref{ss: blue_stragglers}), chemically peculiar stars (\S\ref{ss: chemically_peculiar_stars}), and more general `close' WD + MS binaries (\S\ref{ss: WD + MS}).
See also the bottom right of Figure~\ref{fig:overview_post_mt}.
We subdivide the latter by their observational discovery method (i.e., spectroscopic, astrometric or self-lensing binaries), as these different methods reflect a distinct range of (orbital) parameter space probed.

\subsection{Blue straggler stars \label{ss: blue_stragglers}}
Blue straggler stars are anomalously hot and luminous main-sequence stars that lie above the main-sequence turn-off in color–magnitude diagrams, making them appear younger than the surrounding stellar population. 
There are currently three primary theories for blue-straggler \textit{binary} formation: I) direct stellar collisions during dynamical interactions in dense cluster environments, II) the merger of an inner binary in a triple system, and III) binary mass transfer \citep[see e.g., recent reviews by ][and references therein]{2024arXiv241010314W,2025ARA&A..63..467M}. 
In the binary mass transfer channel, blue stragglers are post–mass transfer accretors, sometimes accompanied by a faint or unseen white dwarf companion. 

In our catalog, we only consider \textit{binary} blue straggler stars with complete orbital solutions.
We include systems from four well-studied open clusters (NGC 2506, NGC 6819, NGC 7789 and NGC 188) with extensive radial-velocity observations that provide kinematic memberships and orbital solutions for the blue straggler populations. 
In particular, we draw on spectroscopic binary orbits from the WIYN Open Cluster Study for NGC~188 \citep{2009AJ....137.3743G}, NGC 6819 \citep{2014AJ....148...38M}, NGC 2506 \citep{2024AJ....168..205L} and NGC~7789 \citep{2020AJ....160..169N}. 
In the case of NGC 188, there are ultraviolet observations of NGC~188 blue stragglers from \citet{2015ApJ...814..163G} that firmly establish the presence of white dwarf companions for at least 1/3 of the blue straggler population, thus confirming their mass-transfer origins. 
Recently, \cite{2026AJ....171..102N} retrieved full spectroscopic-binary orbits for 35 systems in NGC 188. Four of their updated orbits considered the BSS systems WOCS- 4230, 4945, 5020, and 8104, which we also include in our catalog. 
In other clusters, white dwarf companions are not directly detected for these blue stragglers, but considered likely due to wide orbital periods and low binary mass functions.

These examples are among the best studied cluster blue straggler populations. Though blue stragglers are common in star clusters \citep[e.g.,][]{2021A&A...650A..67R}, often they lack orbital information and their formation channel is clearly not known, and we do not provide a full census of all blue stragglers reported in the literature.

We note that there are also other examples of post-mass-transfer systems found in clusters that are related to the blue stragglers and seem to share similar orbital properties. These include evolved counterparts to the blue stragglers known as ``yellow stragglers", and ``blue lurkers" which may be less luminous accretors that blend photometrically with normal main sequence stars \citep{2019ApJ...881...47L}, but we currently do not include those populations in this study.

\subsection{Chemically peculiar stars \label{ss: chemically_peculiar_stars}} 
Accretor stars can be identified through their unusual surface abundances, most notably enhancements of carbon and s-process elements. This includes barium stars, CH stars, carbon- and s-process–enhanced metal-poor (CEMP-s) stars, dwarf carbon (dC) stars, and extrinsic S stars. 
Despite this maze of observational classifications, these groups are closely related: barium, CH, and CEMP-s stars primarily differ by decreasing metallicity, while dC stars represent their lower-mass main-sequence counterparts (overlap with `blue lurkers' in clusters), and extrinsic S stars are the evolved giant analogues lacking technetium. 
They can all be understood as a situation where we observe the accretor star that has gained mass/been chemically polluted by a companion asymptotic giant branch star that underwent thermal pulses and synthesized carbon and s-process elements. 
For a comprehensive overview of these systems, their taxonomy and its historical development, we refer to the review by \citet{2025ARA&A..63..467M} and its supplemental material (see, e.g., their Supplemental Figure 7).

For our catalog, we draw systems from radial-velocity surveys of Barium and related stars, and their white-dwarf companions by \citet{2019A&A...626A.127J} and \citet{2019A&A...626A.128E}. 
The first study focuses on giant stars, and includes both barium and extrinsic S stars, while the second targets barium dwarfs and CH subgiants. 
Together, these studies provide orbital solutions for 82 systems and component-mass estimates available for many of them.


\subsection{Close WD + MS binaries \label{ss: WD + MS}}
White dwarf + main-sequence (WD + MS) binaries are expected to have experienced a mass-transfer episode in their past if they are observed in close orbits today. 
We define ``close" orbits as those with separations $\lesssim 1\,$AU ($P_{\rm orb} \lesssim 1000\,$d), for which Roche lobe overflow is expected to have occurred when the WD progenitor became a giant.
Prior to the Gaia mission and large scale spectroscopic surveys, there existed an inhomogeneous sample of close WD + MS binaries discovered primarily by their radial velocity and photometric variability \citep[e.g.][]{2003A&A...406..305S}. 
Today, the main sources of WD+MS detections are spectroscopic and astrometric binaries, which we consider individually in the following sections. 
Lastly, we discuss systems that have been discovered through self-lensing.

\paragraph{Spectroscopic WD+MS binaries}\label{ss: spectroscopic WDMS}
The Sloan Digital Sky Survey (SDSS) has led to the identification of thousands of WD+MS binary candidates \citep[][]{2006ApJS..162...38A}. For a fraction of these, binary parameters have been obtained from repeated RV measurements \citep[][]{2007MNRAS.382.1377R}, with orbital periods compiled by \citet{2010A&A...520A..86Z} and extended in subsequent studies \citep[e.g.,][]{2007MNRAS.382.1377R, 2012MNRAS.419..817P}. We also note that there have been a few samples of eclipsing WD+MS binaries \citep{2015MNRAS.449.2194P,2026PASP..138c4202S}. 
The majority of these systems host M-type MS stars which are sufficiently dim in the blue optical for the WD to be visible in their spectra. 
To find the missing binaries with more massive (AFGK-type) MS companions, the ``WD binary pathways survey" 
combined optical spectroscopy with GALEX UV photometry and HST follow-up, deriving orbital parameters for systems with significant RV variability \citep{2016MNRAS.463.2125P}. 

Most of the aforementioned systems have short orbital periods $\lesssim 1 \mathrm{d}$, and have been classified as post-common envelope binaries. 
Under this assumption, the common envelope efficiency factor ($\alpha$) required to explain their present orbits can be computed. 
The majority of systems are consistent with $\alpha < 1$, while a single value of $\alpha \simeq 0.2$–$0.3$ can best reproduce most systems at once \citep[note that the precise definition of $\alpha$ varies across the literature, e.g.,][]{2010A&A...520A..86Z, 2011MNRAS.411.2277D, 2010MNRAS.403..179D, 2012MNRAS.419..287D}. 
A notable exception is IK~Peg, which has a relatively long orbital period of 22,days \citep{1993MNRAS.262..277W}, and hosts a massive WD ($\sim1.2\Msun$). 
Its wide orbit implies limited orbital shrinkage during mass transfer, suggesting that any common-envelope phase would require additional energy sources beyond orbital energy \citep[most notably recombination energy e.g.,][]{2010A&A...520A..86Z, 2010MNRAS.403..179D, 2007MNRAS.382.1377R}. \cite{2024MNRAS.52711719Y} further identified five systems hosting ultramassive ($\gtrsim 1.2\,M_{\odot}$) WD candidates around MS stars with orbital periods between $18-49\,$d in the NSS catalog (four spectroscopic, one astrometric), making them IK~Peg analogs. \\


\paragraph{Astrometric WD + MS binaries}\label{ss: astrometric WDMS}
The third data release (DR3) of the Gaia mission introduced the non-single stars (NSS) catalog, containing orbital solutions for over 160,000 astrometric binaries \citep{2022yCat.1357....0G}.
Astrometric measurements are most sensitive to orbital periods between $\sim 100-1000\,$d. For  WD + MS binaries, this roughly corresponds to AU-scale separations, which is significantly wider than previously known binaries while still being smaller than the maximum radii reached by the WD progenitors. 
%

 \citet{2024MNRAS.529.3729S} identified over 3000 high-probability WD+MS binary candidates with astrometric solutions at wide orbital periods. 
The majority of these systems host WDs with masses $\gtrsim 0.6\,M_{\odot}$ which suggest that they originated from AGB progenitors. For these systems, full orbital parameters, including inclinations, are available and have been shown to be broadly reliable \citep{2024PASP..136h4202Y}.
%
\citet{2024MNRAS.529.4840G} furthermore cross-matched WD + FGK MS candidates from spectroscopic surveys to the NSS catalog to confirm their binary nature, creating a sample of 206 systems with orbital periods and WD mass constraints. Note that this sample also includes systems with single-lined spectroscopic orbits from the NSS catalog with periods down to $\sim 1\,$d.

\paragraph{Self-lensing WD + MS binaries}\label{ss: self_lensing_binaries}
A self-lensing binary (SLB) is a binary hosting a compact object which gravitationally lenses the light of its luminous companion during eclipse. The resulting magnification leads to a characteristic peak in the light curve which can be fitted to get binary parameters. 
So far, five WD SLBs have been discovered using data from the Kepler mission. The first of these was KOI-3278, discovered by \citet{2014Sci...344..275K}. This system has an orbital period of 88 days, longer than most known post-CE systems at the time, posing similar challenges to IK~Peg in understanding its evolutionary history. 
Since then, four more candidates have been confirmed \citep{2018AJ....155..144K, 2020IAUS..357..215M, 2024PASP..136g4201Y}, with longer periods than KOI-3278 ($\sim 420-730\,$d). 
They lie in a similar region of the orbital period-WD mass parameter space as the Gaia astrometric binaries \citep{2024MNRAS.529.3729S}, meaning that their mass transfer histories are similarly uncertain. Considering eclipse probabilities, the five known SLBs suggest an occurrence rate of $\sim 1\%$ of all solar-type stars hosting WD companions in AU-scale orbits \citep{2024PASP..136g4201Y}. This value is consistent with an independent constraint from the forward-modeling of the astrometric sample by \citet{2025PASP..137j4205Y}. 

Our WD+MS sample comprises the majority of our catalog, with \allWDMS \, systems drawn from several large observational catalogs in which the presence of a white dwarf companion is inferred, though not always directly confirmed. 
The bulk of our catalog derives from the SDSS spectroscopic WDMS catalogs, including the DR7 compilation of 2248 binaries \citep{2012MNRAS.419..806R} and follow-up studies by \citep{2010A&A...520A..86Z}. 
All data from WD Binary Pathways VI \citep{2022MNRAS.512.1843H} and X \citep{2024MNRAS.529.4840G} is also included.
Systems from these studies are labeled `Spectroscopic WD + MS' in our catalog.
We further add the WD+MS binaries identified through Gaia DR3 astrometric orbits \citep{2024MNRAS.529.3729S} and the UV-excess–selected samples with Gaia orbital solutions from \cite{2024MNRAS.529.4840G}. 
Systems from these sources are labeled `Astrometric WD + MS' in the catalog.
Lastly, we also ensure that all systems from the post-common envelope catalog presented by \citet{2021ApJ...920...86K} are included, and these are classified simply as `WD + MS' in our catalog.

\section{Binaries containing a neutron star or black hole}
\label{sec: NSBHbinaries}
Finally, we discuss binary systems that show evidence for a NS or BH companion, i.e., the C.O.\ binary category in Table \ref{tab:obs_classes} and the top right of Figure~\ref{fig:overview_post_mt}.
We begin with systems in which the presence of a NS or BH is inferred from NS pulsations (\S\ref{ss: pulsar_binaries}) or accretion signatures, i.e., high-mass X-ray binaries (\S\ref{ss: XRB}). We then consider dormant compact-object binaries, where the companion is inferred indirectly, typically because the inferred mass of the unseen companion exceeds the maximum for a WD (or NS).
Specifically, we include systems where the compact object is inferred based on spectroscopic (\S\ref{ss: spectroscopic_CO}), or astrometric (\S\ref{ss: Gaia_CO}) variability of its luminous companion.

\subsection{Pulsar binaries with alive companions \label{ss: pulsar_binaries}}
Pulsar binaries host a NS detected through radio pulsations. Their orbital parameters can be determined through pulsar timing, which constrains the binary motion via variations in pulse arrival times \citep[e.g.,][]{1989ApJ...345..434T}. Since massive stars are expected to interact with their companion over their lifetime, pulsar binaries are expected to have undergone at least one phase of mass transfer \citep[e.g.,][and references therein]{2017ApJ...846..170T}. 

We restrict the sample in our catalog to young pulsars with non-degenerate companions. In these systems, the NS progenitors were the initially more massive component. There was likely a period of stable mass transfer in which the progenitor overflowed its Roche lobe, losing its hydrogen envelope to its main sequence companion and widening the binary orbit \citep[e.g.,][]{2017ApJ...846..170T}. The resulting stripped star then exploded as a supernova (SN), leaving behind the young pulsar. The present-day orbital periods and eccentricities of these systems can provide constraints on the natal kicks imparted to NSs due to mass loss and SN asymmetry \citep[e.g.,][]{1994Natur.369..127L}.

The majority of pulsars detected in binaries are ``recycled'' pulsars, i.e., NSs that have been spun up by their accretion from their companion \citep[see e.g., Fig. 1.2 in ][]{2023pbse.book.....T}. Many of these are also identified as millisecond pulsars \citep[MSPs, e.g.,][]{1991PhR...203....1B}. 
Since recycled pulsar binaries have likely undergone more than one phase of mass transfer, we do not include them in our catalog (see Appendix~\ref{sec: not incl} for discussion).

We include the following seven young pulsar–OB star binaries in wide orbits discovered through pulsar timing:
PSR J0045-7319 \citep{1995ApJ...447L.117B, 1996Natur.381..584K}, PSR J1740-3052 \citep{2001MNRAS.325..979S, 2011MNRAS.412L..63B, 2012MNRAS.425.2378M}, PSR J1638-4725 \citep{2006MNRAS.372..777L}, PSR B1259-63 \citep{2014MNRAS.437.3255S}, PSR J2032+4127 \citep{2017MNRAS.464.1211H}, PSR J2108+4516 \citep{2023ApJ...943...57A}, and PSR J0210+5845 \citep{2024A&A...682A.178V}.
On top, we also include two young pulsar binaries with roughly solar-mass companions on eccentric orbits (PSR B1820-11; \citealt{1989Natur.340..367L}, and PSR J1954+2529; \citealt{2022ApJ...924..135P}).
In the latter systems, the luminous companion remains unidentified and may be a white dwarf rather than a main-sequence star. In all cases, the mass function is well measured, but companion masses must be inferred by assuming a typical NS mass and orbital inclination.
%


\subsection{X-ray binaries \label{ss: XRB}}
X-ray binaries (XRBs) host an accreting compact object which is most commonly a NS, and more rarely a BH. 
XRBs are further classified based on the nature of the donor star, distinguishing low-mass and high-mass X-ray binaries.

\paragraph{Low-Mass X-ray Binaries}
The majority of known XRBs are low-mass XRBs (LMXRB), in which the compact object accretes though (long-term) Roche-lobe overflow from its companion. 
LMXBs are notoriously hard to form. 
However, the most commonly proposed formation channels, such as post–common-envelope evolution \citep{1996ApJ...458..301K, 1998ApJ...493..351K, 1998ApJ...493..368K}, triple-driven evolution \citep[e.g.,][]{2016ApJ...822L..24N,2025ApJ...983..115S}, 
or dynamical formation in dense clusters \citep[e.g.,][]{1987IAUS..125..187V}, either assume the LMXB phase is the second mass-transfer episode or invoke additional dynamical processes to bring the donor and compact object into contact. 
In either case, we do not expect them to cleanly probe post–mass-transfer properties, and we therefore exclude them 
(see Appendix~\ref{sec: not incl} for discussion).

\paragraph{High-Mass X-ray Binaries}
In high-mass X-ray binaries (HMXBs), the compact object accretes from a massive stellar companion. 
HMXRB can be further subdivided into Be X-ray binaries (typically in wide, eccentric orbits where the compact object accretes from the decretion disk of its Be-type companion), and supergiant X-ray binaries (where accretion proceeds primarily through the stellar wind of an evolved OB supergiant), with additional (rarer) subclasses such as supergiant fast X-ray transients (SFXTs), or systems containing sgB[e] or Wolf–Rayet donors \citep[e.g.,][]{2023A&A...671A.149F}. 
While BeXRBs are typically detached and cleanly probe the post–first-mass-transfer state, wind-fed XRBs can be edge cases between the `first' and `second' mass-transfer phases. 
We include both, as together they sample a range of post-interaction separations.

We include all 162 systems from the most recent version (v2024-08)\footnote{\url{https://binary-revolution.github.io/HMXBwebcat/}} of the HMXB catalog of \citet{2023A&A...671A.149F}. 
Subclassifications are recorded in the \texttt{Notes} field where available. 
We adopt the compact-object classification provided by the catalog and additionally set \texttt{evol\_type\_2} to NS or BH when a measured compact-object mass is available and satisfies $M \leq 2.5\,\Msun$ or $M > 2.5\,\Msun$, respectively.



\subsection*{Dormant compact-object binaries \label{ss: dormant_CO}}
Dormant compact-object binaries are systems in which a BH or NS is present, but neither accreting at observable levels nor detectable as a radio pulsar. 
In such systems, the compact object is revealed only indirectly, through the spectroscopic, astrometric, or photometric variability of its luminous companion.
To date, only spectroscopic and astrometric variability have led to successful detections of dormant compact-object binaries. 


\subsection{Spectroscopic compact-object binaries \label{ss: spectroscopic_CO}}  
Spectroscopic searches for dormant BHs and NSs date back to before the discovery of the first X-ray binaries \citep{1966SvA....10..251G, 1969ApJ...156.1013T}. Despite decades of effort, such systems are intrinsically rare and difficult to identify, while false positives can arise from stripped-star binaries (see Sections~\S\ref{ss: intermediat M stars} and \S\ref{ss: EL CVn}). 
Only recently have extensive spectroscopic surveys successfully uncovered dormant BHs, and we include the following four currently confirmed systems in our catalog: NGC 3201 $\#$12560 \citep{2019A&A...632A...3G}, VFTS 243 \citep{2022NatAs...6.1085S}, G3425 \citep{2024NatAs...8.1583W}, and HD 130298 \citep{2022A&A...664A.159M}.

Furthermore, several candidates for dormant NSs have been identified via radial velocity follow-up, although in all cases the unseen companion could instead be a massive WD.
Specifically, we include the following of such systems in our catalog:
2XMM J1255 \citep{2022MNRAS.517.4005M}, 2MASS J06163552 \citep{2022ApJ...940..165Y}, LAMOST J2354 \citep{2023SCPMA..6629512Z,2025OJAp....8E..61T}, LAMOST J1123 \citep{2022NatAs...6.1203Y}, 56 UMa \citep{2023A&A...670L..14E}, 2MASS J1527 \citep{2023ApJ...944L...4L,2024ApJ...961L..48Z}, J0606+2132 \citep{2024ApJ...977..245Z}, G3431, G8441, and G4031 \citep{2024ApJ...964..101Z}.




\subsection{Astrometric compact-object binaries \label{ss: Gaia_CO} } 
In contrast to most binary-detection methods described above, \textit{Gaia} \citep{2016A&A...595A...1G} astrometry is most sensitive to binaries with \textit{wide} orbital separations. 
Investigation of Gaia orbital solutions with high mass functions has led to the discovery and characterization of three nearby dormant BHs 
\citep[`Gaia BHs',][]{2023MNRAS.518.1057E,2023AJ....166....6C,2024PASP..136a4202N, 2023ApJ...946...79T,2023MNRAS.521.4323E,2024A&A...686L...2G}, 
and a population of twenty-one dormant NSs \citep[`Gaia NSs'][]{2024OJAp....7E..27E,2024OJAp....7E..58E} in au-scale orbits.
Population synthesis studies calibrated to the DR3 detections predict discovery of dozens more Gaia BHs and NSs in future data releases \citep[e.g.,][]{2025PASP..137d4202N,2025arXiv250821805C}.

Gaia BHs and NSs are likely post–mass transfer binaries, since their current orbital separations are significantly smaller than the predicted maximum radii of the BH or NS progenitors \citep[e.g.,][]{2022ApJ...931..107C}. 
Yet, standard isolated binary evolution fails to reproduce the observed au-scale orbits, as the extreme mass ratios at interaction should result in unstable mass transfer. Any systems that survive common-envelope evolution are expected to emerge with much tighter orbits.
As a result, a range of alternative formation channels have been proposed, including modified 
isolated-binary evolution \citep[e.g.,][]{2024A&A...692A.141K,2024MNRAS.535L..44G, 2024MNRAS.535.3577K,2024A&A...690A.144I,2025arXiv251110728O}, 
hierarchical triple evolution \citep[e.g.,][]{2024PASP..136a4202N, 2024ApJ...964...83G, 2024ApJ...975L...8L, 2025OJAp....8E..79T}, and dynamical assembly in star clusters \citep[e.g.,][]{2023MNRAS.526..740R, 2024MNRAS.527.4031T, 2024ApJ...965...22D, 2024A&A...688L...2M}. 
To date, their origin remains an open question, making it especially valuable to study their properties in the context of other post‑mass-transfer systems.

Recently, oMEGACat BH-2 was discovered through astrometric analysis of archival and new Hubble Space Telescope (HST) data \cite{2026arXiv260618350W}.
This system is remarkable because it hosts a stellar-mass BH in a low-metallicity globular cluster, with a mass of $4.46^{+1.22}_{-1.01}\Msun$, placing it in the putative NS--BH mass gap \citep[first discussed in][]{1998ApJ...499..367B,2010ApJ...725.1918O,2011ApJ...741..103F}. 
It also has the longest orbital period of any known BH binary, making it unlikely that its present-day orbital properties were significantly shaped by mass transfer. Nevertheless, we include it in our catalog but flag it as a special case.


\section{Exploring the properties of post-mass transfer systems} \label{sec:correlations}


In this section, we explore the correlations between various post-mass transfer properties of the systems in our catalog. We focus on the relationships between the mass ratios, orbital periods, and other key parameters of the binaries, aiming to identify any trends that may provide insights into the evolution of these systems.

Interactive versions of these plots are available at \url{https://lvanson.github.io/post_mt_relations.github.io/post_mt_catalog/#plots}.

\subsection{Period-eccentricity relations \label{ss: P-e diagram}}

\begin{figure*}
\centering
\includegraphics[width=0.9\textwidth,trim=0.cm 0mm 0cm 0cm,clip]{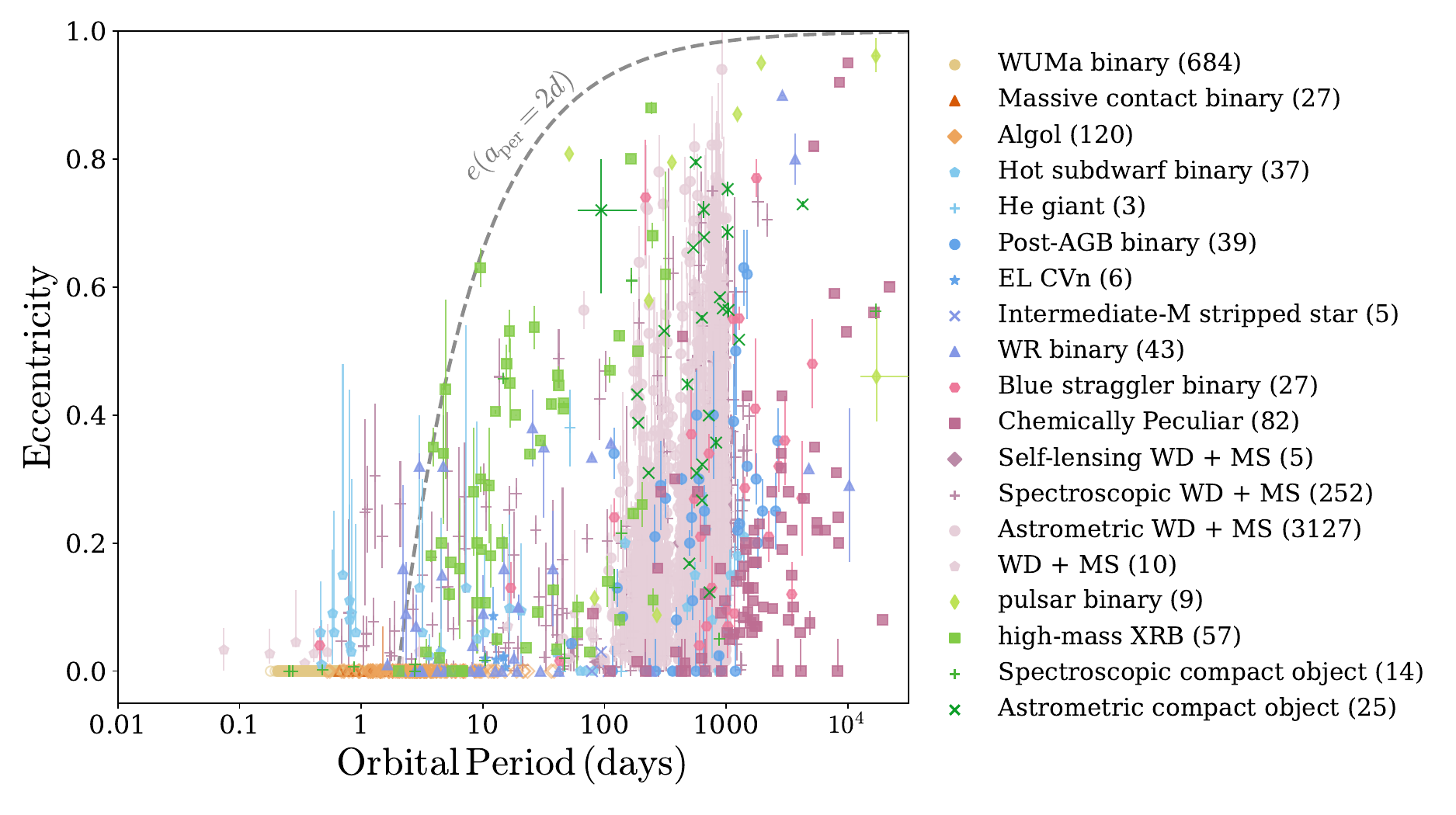} 

\includegraphics[width=0.54\textwidth]{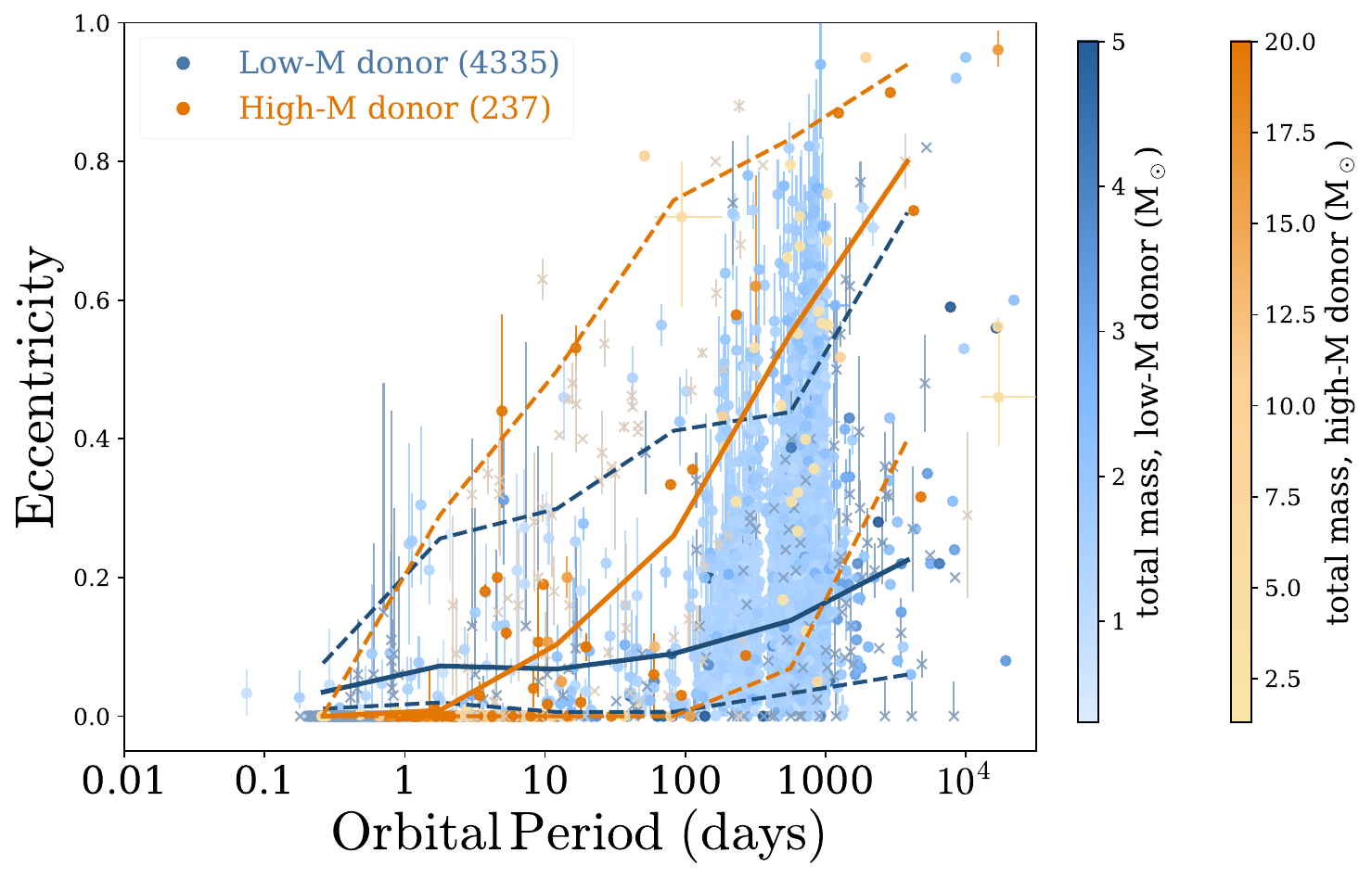} 
\includegraphics[width=0.45\textwidth]{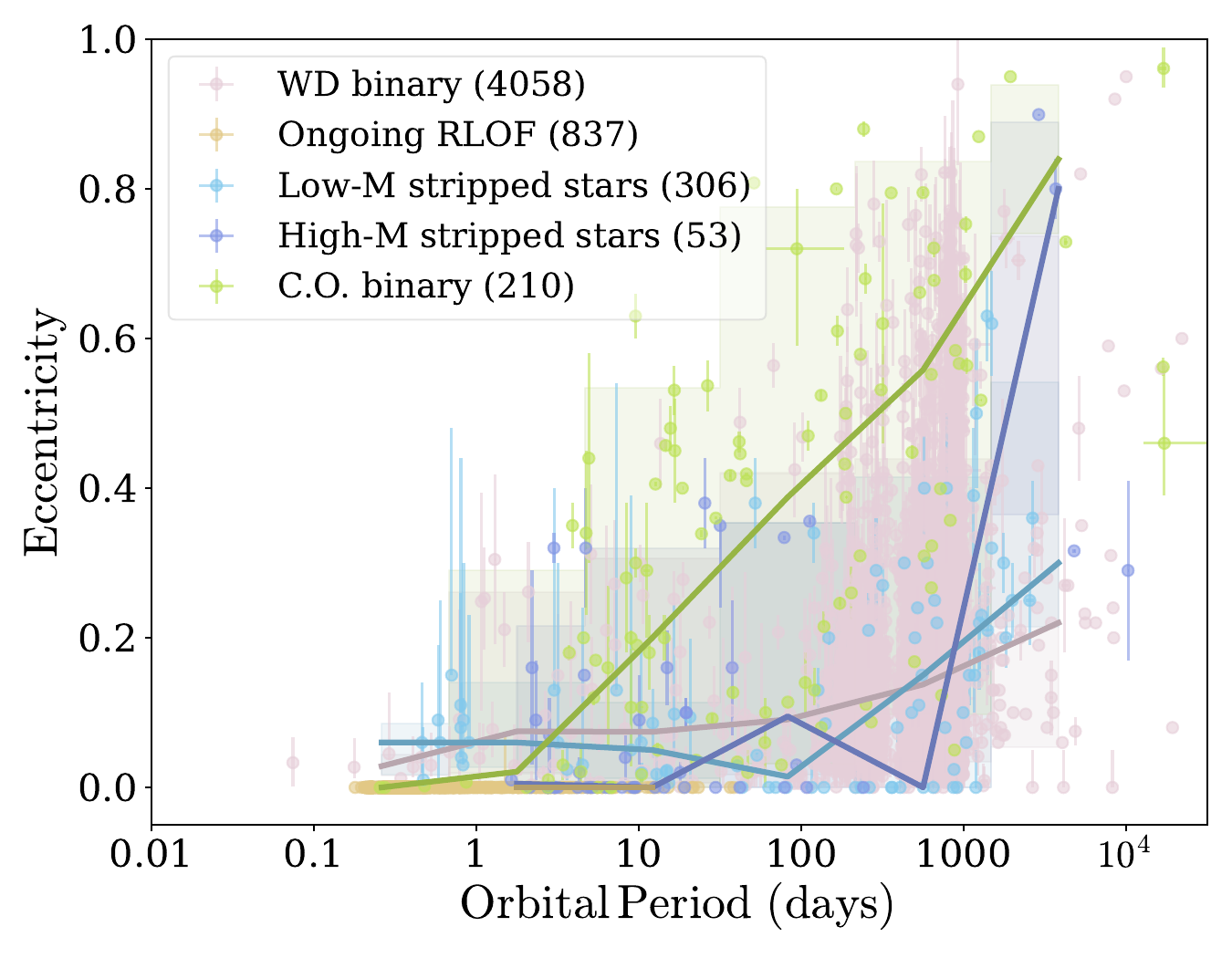} 

\caption{Variations of the $P$--$e$ diagram colored by I) system class (top panel), II) total mass (bottom left), and III) the evolutionary category (Bottom Right). The number of systems for each entry is indicated between parentheses in the legend.
The top panel also shows equation \ref{eq: emax} as a dashed line.
We show the median and $90\%$ eccentricities intervals in the bottom two plots with thick lines and dashed lines/shaded regions, respectively. Assumed or inferred values are not included in the median calculations, and are shown as empty symbols in the top panel.
An interactive version of the top panel 
\protect\href{https://binary-observations.github.io/post_mt_catalog/\#plots}{is available here.}
While non-zero eccentricities are present across all periods, longer-period systems show a broader eccentricity distribution than shorter-period systems. We find a hint of mass dependence as systems from presumably low-mass donors have lower median eccentricities than those from high-mass donors.
 The Code used to generate this plot \protect\href{https://github.com/binary-observations/post_mt_catalog/blob/master/code/data_analysis/P_e_Figures.py}{can be found here.} 
 }
\label{fig:P_e_diagrams}
\end{figure*}

The orbital period–eccentricity ($e-\log P$) diagram has long been of interest in the context of binary mass transfer. 
Theoretically, tidal interactions are expected to be efficient enough to circularize systems prior to RLOF \citep{1975A&A....41..329Z,1977A&A....57..383Z,1995A&A...296..709V}, implying that mass transfer should occur in circular orbits.
We note that \cite{2024A&A...681L...1S} show that implementations of the original \cite{1977A&A....57..383Z} prescriptions often deviate from their original formulation, an inconsistency that can alter the tidal evolution path toward synchronization.
This circularization assumption forms the foundation that allows us to adopt Roche-lobe formalism for mass transfer \citep{1983ApJ...268..368E,1988A&A...202...93R,1990A&A...236..385K}, which is implemented across almost all modern binary evolution codes \citep[e.g.,][and all codes derived from these works]{1971MNRAS.151..351E,2002MNRAS.329..897H,2015ApJS..220...15P}. 
We thus expect short-period (presumably post-mt) systems to be near-completely circularised, while longer-period systems may retain a broader eccentricity distribution if they have not undergone mass transfer.
Yet, it has long been noted that systems presumed to have undergone mass transfer are not as strictly circular as expected from theory. 
Early evidence came from barium stars \citep[][]{1998A&A...332..877J}, whose period-eccentricity diagram shows lower maximum eccentricities than normal giants at the same orbital period.
Yet even in this population, many long-period systems retain significant eccentricities. 
Subsequent studies have shown that this behaviour is not unique, but instead occurs across a wide range of post-interaction populations, including 
post-AGB binaries \citep[e.g.][]{2009A&A...505.1221V,2018A&A...620A..85O,2025A&A...703A.294M},
hot subdwarf (sdB/sdOB) binaries \citep[e.g.][]{2015A&A...579A..49V,2017A&A...605A.109V,2020A&A...641A.163V,2022A&A...658A.122M}, 
CH and CEMP stars \citep[e.g.][]{2019A&A...623A.128H}, 
and blue stragglers \citep[e.g.][]{2009Natur.462.1032M,2011Natur.478..356G}.

By compiling the largest catalog of post-mass-transfer systems to date, we can assess the prevalence of eccentric orbits across different types of post-mass transfer systems, as well as search for correlations between the eccentricity and evolutionary class or system mass.

In the top panel of Figure~\ref{fig:P_e_diagrams}, we show the $e-\log P$ distribution for all systems with measured $\log P$ and $e$ (including uncertainties where available), color-coded by system class. 
Tidal dissipation is generally assumed to circularize binaries with $P \lesssim 2\,\mathrm{d}$ on timescales short compared to the stellar lifetime \citep[e.g.,][]{1989A&A...220..112Z, 2005ApJ...620..970M}. 
This motivates the dashed line, which shows Equation~3 from \citet{2017ApJS..230...15M}: 

\begin{equation}
e_{\max} = 1 - (P/2\,\mathrm{d})^{-2/3}.
\label{eq: emax}
\end{equation}

I.e., showing where the periastron separation would equal the semi-major axis of a circular binary at the 2-day circularization period. Systems above this line are commonly assumed to circularize rapidly due tides.
We find that non-zero eccentricities are present across the full period range, with no clear regime of exclusively circular orbits. 
The only circular populations are the observed contact binaries and Algol systems, but we stress that this is largely a modelling choice: even though we do not expect high eccentricities in these systems, the light-curve models used to fit these systems almost always \textit{assume} $e=0$ a priori \citep[e.g.,][]{2025A&A...695A.223V,2025A&A...695A.109F}, so the data are never given the opportunity to constrain the eccentricity. 
We therefore exclude them from the population-level medians below.
Conversely, we further caution that a measured low eccentricity does not per se imply a truly circular orbit: because $e \geq 0$, measurement noise produces a one-sided distribution that biases small recovered values upward, and small catalogued eccentricities ($e \lesssim 0.05$) are often statistically consistent with $\varepsilon = 0$ \citep[e.g.,][]{1971AJ.....76..544L,2005A&A...439..663L,2013A&A...551A..47L}. 
Regardless of these individual-system caveats, a robust global trend emerges from our sample: at short orbital periods, the eccentricity distribution is narrow with a low median. 
Toward longer periods, both the median eccentricity and the width of the distribution increase.
To investigate the median of the $e(\log P)$ relation, we first subdivide the catalog by both mass and category (see Table~\ref{tab:obs_classes}). This is necessary because the catalog is heavily dominated by low-mass sources (in particular, white dwarf binaries) and contains a diverse collection of systems. Partitioning the data into more homogeneous subsamples enables a more physically meaningful comparison of the $e(\log P)$ relation across different post-mass transfer populations.

\paragraph{Is the $e(\log P)$ relation mass dependent?}
In the bottom-left panel of Fig.~\ref{fig:P_e_diagrams}, we divide the systems into ``presumably low-mass donors'' and ``presumably high-mass donors.'' 
For this purpose, we assume the following sytem classes descend from low-mass donor stars: 
Hot subdwarf binary, He giant, Post-AGB binary, EL CVn, Blue straggler binary, Chemically Peculiar, and all WD containing binaries. 
Similarly, we assume the following system classes descend from high-mass donor stars: 
Intermediate-M stripped star, WR binary, pulsar binary, high-mass XRB, Spectroscopic compact object, Astrometric compact object. \footnote{We caution that some of the spectroscopic compact-object binaries, listed as having a NS companion, might instead have a WD companion (see \S \ref{ss: spectroscopic_CO}). Particularly LAMOST J2354 \citep{2023SCPMA..6629512Z,2025OJAp....8E..61T}, 2MASS J1527 \citep{2023ApJ...944L...4L,2024ApJ...961L..48Z}, and G4031 \citep{2024ApJ...964..101Z}.} 
%
We apply this donor-based classification rather than dividing by observed mass because system classes more directly reflect the donor’s evolutionary history.
Nonetheless, we also show total system mass as a color scale in both panels. 
We note that our definition of low-mass donor systems span $M_{\mathrm{tot}} \approx 0.3$–$18.6\,\Msun$, overlaps in total mass with the high-mass donor systems, which span $M_{\mathrm{tot}} \approx 1.3$–$92.2\,\Msun$. 

We find that the median eccentricity is systematically higher for high-mass donor systems. For low-mass donors, the median eccentricity increases from $\sim 0.05$ at $\log P \sim -0.5$ to $\sim 0.2$ at $\log P \sim 3.5$. In contrast, for high-mass donors it rises from $\sim 0$ to $\sim 0.7$ over the same period range.
Interestingly, this is consistent with the conclusion from \cite{2024OJAp....7E..58E} who compared the Period-eccentricity relation and found that NS+MS systems have significantly higher eccentricities than WD+MS systems at fixed period.
We caution that the number of high-mass donor systems is much smaller, making these medians less robust. Our binning is chosen such that most $\log P$ bins contain $\gtrsim 30$ systems, although at $\log P \gtrsim 3.2$ only $\sim 10$ high-mass systems remain. Similarly, below $\log P \lesssim -0.1$, both subsamples contain fewer than $\sim 10$ systems per bin, and the inferred trends become uncertain.

\paragraph{Does the $e(\log P)$ relation depend on evolutionary category?}
We next examine the  $e(\log P)$ relation by subdividing the catalog by evolutionary category, as shown in the bottom-right panel of Fig.~\ref{fig:P_e_diagrams}.
The median eccentricities of the `WD binary' (pink, right panel) and `low-mass stripped' (blue, left panel) sets are very similar, consistent with low-mass stripped stars being the progenitors of (a subset of) low-mass WDs.
Low-mass stripped stars broadly follow the same $e(\log P)$ trend, with a possible dip toward lower eccentricities around $\log P \sim 2$.
The`On-going RLOF' category is shown, but since almost all eccentricities in this category are \textit{assumed} to be zero, we do not plot their median $e(\log P)$ relation, as it has not constraining power.
High-mass stripped stars exhibit generally low median eccentricities across most orbital periods, with two apparent exceptions: at $\log P \sim 2$, where the median eccentricity is approximately 0.1, and at $\log P \sim 3$, where the median eccentricity rises to about 0.8. 
We do not, however, place significant confidence in these trends because the sample sizes in the relevant period bins are small: there are only 8, 1, and 3 systems in the $\log P$ intervals (1.5-2.333], (2.33-3.17], and (3.17-4.0], respectively.

In contrast, compact-object binaries exhibit systematically higher eccentricities than the other categories. Their median eccentricity is already $\sim 0.2$ at $\log P \sim 1$ and increases steadily to $\sim 0.8$ at $\log P \sim 3.5$. 
We note that the compact-object binaries are responsible for the elevated median eccentricities observed for the high-mass donor categories in the bottom-left panel of Figure~\ref{fig:P_e_diagrams}.
It is somewhat debatable whether XRBs are truly in a post–first mass-transfer phase as opposed to already undergoing a second mass-transfer phase (see \S \ref{ss: XRB}). 
To correct for this, we have also examined the median $e(\log P)$ relation while excluding the XRBs, and we find no significant difference. 
This hints towards both categories sharing a similar evolutionary origin and are consistent with being post–first mass-transfer systems.

\subsubsection*{Potential solutions to the non-zero eccentricities}
Two broad categories of solutions have been proposed: either the orbits do not fully circularise during mass transfer, or some eccentricity-pumping mechanism operates during or after mass transfer to restore eccentricity.

Several mechanisms have been proposed to pump eccentricity during or shortly after the mass transfer phase. 
First, tidally-enhanced wind mass loss can introduce a phase-dependent asymmetry in mass loss from the donor that counteracts circularization. 
The underlying prescription, in which tides enhance the donor's wind, was originally introduced by \citet{1988MNRAS.231..823T} in a different context \citep[see also][]{2006epbm.book.....E}; its role as an eccentricity-pumping mechanism was subsequently discussed by \citet{2000A&A...357..557S,2008A&A...480..797B,2014A&A...565A..57S}.

Second, phase-dependent mass loss during RLOF on eccentric orbits (where mass transfer is concentrated near periastron and low near apastron) can generate eccentricity that opposes tidal circularization 
\citep{2007ApJ...667.1170S,2009ApJ...702.1387S,2016ApJ...825...70D,2019ApJ...872..119H}. 
Recently, \cite{2025ApJ...983...39R}  used \MESA to implement the analytic expressions for secular rates of change of the orbital semi major axis and eccentricity, assuming a delta function mass transfer at periapse. 
They showed that eccentric mass transfer onto compact objects can naturally preserve significant eccentricity post-mass-transfer, without invoking instantaneous circularization. 
\cite{2019ApJ...872..119H} present an analytic model to describe the secular orbital evolution of binaries undergoing conservative mass transfer by implementing a mass transfer rate that is a smoothly varying function of orbital phase, rather than a delta function centered at periapsis.
\cite{2026A&A...706A..79P,2026A&A...706A.357P} expanded upon this model to include both conservative and non-conservative mass transfer valid at arbitrary eccentricity. They find a broader parameter space for eccentricity pumping than previous models, and show that eccentric mass transfer naturally predicts higher eccentricities at longer orbital periods.
A related mechanism, the grazing envelope evolution, in which enhanced mass loss during periastron passages counteracts tidal circularisation, was shown by \cite{2018MNRAS.480.3195K} to maintain high eccentricities during the AGB-to-post-AGB transition.

Third, interactions with a circumbinary  disk can pump eccentricity through resonant torques that transfer angular momentum from the binary orbit to the disk \citep[e.g.,][]{2013A&A...551A..50D,2015A&A...579A..49V}.
This mechanism is observationally motivated by the ubiquitous presence of stable dusty circumbinary disks around post-AGB binaries \citep[e.g.,][]{2006A&A...448..641D,2018A&A...620A..85O}. 
\cite{2020A&A...642A.234O} found that circumbinary disk interactions alone were insufficient to explain the high eccentricities in their sample of depleted post-AGB binaries. 
More recently, Moltzer (in prep) argued that the circumbinary disk mechanism is responsible for boosting eccentricities in post-AGB and post-RGB systems, but noted that some initial seed eccentricity is required; if a system begins perfectly circular, disk-binary resonances are ineffective at pumping eccentricity. 
In this picture, a mechanism such as eccentric mass transfer could provide the initial eccentricity ($e \lesssim 0.1$--$0.2$), which the circumbinary disk subsequently amplifies. 

\cite{2015A&A...579A..49V} tested all three mechanisms (tidally-enhanced wind mass-loss, phase-dependent RLOF on eccentric orbits and the interaction between a circumbinary disk and the binary) using \MESA and found that a combination of phase-dependent RLOF and circumbinary disk interactions was needed to cover the observed period-eccentricity space of long-period sdOB binaries, though they could not reproduce the trend of increasing eccentricity at longer periods.
Additional mechanisms include perturbations from tertiary companions via Lidov-Kozai oscillations \citep[e.g.,][]{2016ARA&A..54..441N,2016ComAC...3....6T,2020A&A...640A..16T}, the retention of some initial eccentricity through the common envelope phase itself, as found in several hydrodynamical simulations \citep{2021MNRAS.507.2659G,2020A&A...644A..60S,2024A&A...683A..65B}, and 
white dwarf natal kicks \citep{2010A&A...523A..10I}.
E.g., \cite{2025arXiv250821805C} found that WD natal kicks as high as 15\,km/s are needed to reproduce the observed eccentricities of WD + luminous-companion systems.

The inability of any single mechanism to reproduce the full range of observed eccentricities across post-interaction binaries suggests that multiple processes may operate simultaneously.
Our comprehensive catalog is intended to help disentangle these contributions. Any successful explanation, whether invoking just one, or several mechanisms, should provide an explanation for the full population of post-mass-transfer systems in a consistent way.

The fact that the rise of $e$ with $\log P$ is shared across nearly all system classes argues for mechanisms that act generically during mass transfer, such as phase-dependent RLOF \citep{2025ApJ...983...39R,2026A&A...706A..79P,2026A&A...706A.357P}, rather than mechanisms requiring channel-specific conditions. 
The elevated eccentricities of compact-object binaries further suggest that their natal kicks contribute on top of this, motivating future comparison of our catalog with population-synthesis predictions that include both effects.

\subsection{Post mass-transfer periods \label{ss: period distribution}}

\begin{figure}
\centering
\includegraphics[width=0.45\textwidth]{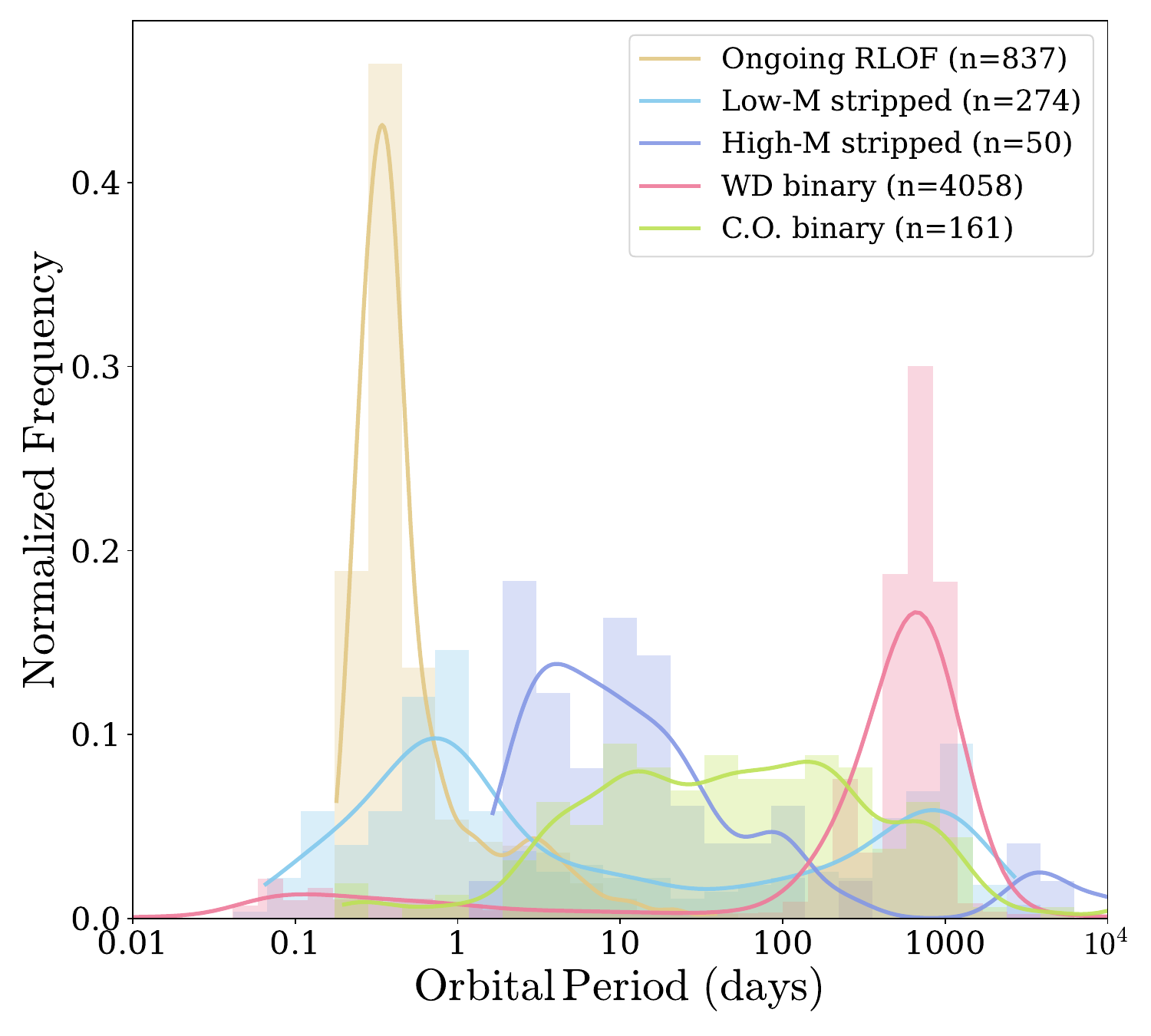} 
\caption{Observed period distribution normalized for different categories of post-mass transfer binaries. Note that this distribution is not completeness-corrected due to the wide range of selection effects. 
Period distributions are broad and strongly overlapping across post-mass-transfer categories. 
\href{https://github.com/binary-observations/post_mt_catalog/blob/master/code/data_analysis/P_dist_Figure.py}{Code used} to generate this plot.}
\label{fig:P dist}
\end{figure}

The orbital period distribution of post-interaction binaries is expected to reflect the distinct evolutionary pathways through which they form. In particular, common-envelope evolution is predicted to produce short-period systems, while stable mass transfer typically leads to wider orbits \citep[e.g.,][]{1984ApJ...277..355W,1976IAUS...73...75P,1997MNRAS.291..732T,1997A&A...327..620S}. 
Systems that have not undergone significant interaction are expected to populate the longest periods as systems with sufficiently wide orbits avoid RLOF altogether \citep[e.g.,][]{2017ApJS..230...15M}.
These channels therefore suggest a qualitative separation in period space:
(i) short-period systems dominated by post-CE binaries,
(ii) intermediate periods populated by systems that underwent stable mass transfer, and
(iii) long-period systems corresponding to non-interacting or weakly interacting binaries.

To investigate this in our sample, we plot the observed period distributions for systems subdivided by evolutionary category in Figure~\ref{fig:P dist}.  
This distribution is not completeness-corrected, and we expect substantial selection effects for different categories. 
Nevertheless, most categories span a very broad range of periods (from about $1$ to $10^4$ days) and show significant overlap.
This suggests that the separation between formation channels may be less distinct in practice, either because the outcomes of stable mass transfer and CE evolution overlap, or because observational biases blur intrinsic features in the distribution.
The only category clearly confined to short periods is `Ongoing RLOF' ($P \lesssim 10$ days). 
This reflects the fact that the time a system spends in RLOF, is short except during case~A mass transfer, making longer-period systems statistically unlikely to be caught mid-transfer.
Taken at face value, the data show no clear separation in period space among post-mass-transfer categories.

\subsubsection*{Post CE or post stable mass transfer? \label{ss: post-CE or stable}}
To investigate this further, we can attempt to break down the systems into those that are more likely to be post-CE and those that are more likely to be post-stable-mt.
In our catalog, we on purpose don't discriminate our post-mass-transfer systems by being potentially post-CE or post-stable-mt.  
Yet we can still guesstimate a bit which system classes are more likely to be post-CE or post-stable-mt based on expectations from the literature, and compare this to our observed period distribution.

Several classes of post-mass-transfer systems are commonly associated with stable mass transfer in the literature.
EL CVn systems are typically interpreted as remnants of stable mass transfer \citep[e.g.,][]{2017MNRAS.467.1874C}.
Wide sdOB binaries are also commonly associated with stable mass transfer, whereas close sdOB+WD and sdOB+MS systems are more often attributed to common-envelope evolution \citep[e.g.,][]{2019MNRAS.482.4592V}.
Generally, mass transfer from high-mass donors is expected to be preferentially stable because these stars have radiative envelopes \cite{1987ApJ...318..794H}, and recent work supports this trend even for giant (more convective) donors \citep[e.g.,][]{2015MNRAS.449.4415P,2017MNRAS.465.2092P,2021A&A...650A.107M}.
Under this assumption, high-mass stripped and compact-object binaries are plausible post-stable-mass-transfer candidates. 
We find that sdOB binaries in our catalog span the full period range, with no clear separation between close and wide systems.
Moreover, many EL CVn systems lie at short periods ($P \sim 1$ day) that are often associated with post-CE separations.
High-mass stripped stars are most abundant between $P \sim 1$--$1000$ days, with a few systems at very wide periods ($P \gtrsim 1000$ days). Compact-object binaries appear in our catalog across a pretty flat distribution between $P \sim 1-1000$ days. 
If these classes are indeed predominantly post-stable-mass-transfer systems, their period distribution is broad and overlaps substantially with that expected for post-CE systems.

\paragraph{WD + MS as post common envelope binaries}
In general, WD+MS binaries are expected to comprise of wide binaries that evolved like single stars, i.e. without interaction ($Porb \gtrsim 3000 $ days), and post-common-envelope binaries (PCEBs, Porb $\lesssim 100$ days) that underwent a common-envelope phase during their evolution \citep[e.g.,][]{2011A&A...536A..43N}.
Classical theory predicts mass transfer from AGB donors are likely dynamically unstable, and leads to a common-envelope, which is expected to produce short post–mass-transfer periods. This is in contrast with observations: 
The majority of the systems identified by \cite{2024MNRAS.529.3729S} host WDs with masses $\gtrsim 0.6\,M_{\odot}$ which suggest that they originated from AGB progenitors. For these systems, full orbital parameters, including inclinations, are available and have been shown to be broadly reliable \citep{2024PASP..136h4202Y}.
The evolutionary history of these systems remain uncertain. 
Recent studies using detailed 1D models of WD progenitors show that wide post–common-envelope binaries can form without invoking extra energy (e.g. from recombination) if the common-envelope phase begins during a thermal pulse of an AGB donor, when the envelope is weakly bound \citep{2024A&A...686A..61B, 2024PASP..136h4202Y}. This channel is particularly relevant for IK~Peg–like systems hosting ultramassive WDs, whose progenitors likely experienced unstable mass transfer due to extreme mass ratios (though the possibility of merger products complicates this interpretation).
Meanwhile, recent theoretical work show that giant donors may be significantly more stable than previously expected \citep{2020ApJ...899..132G, 2023A&A...669A..45T}, and population modeling under this assumption has been able to reconstruct the key features of the \citet{2024MNRAS.529.3729S} sample \citep{2024PASP..136h4202Y}. Alternative scenarios, such as grazing envelope evolution, have also been proposed \citep{2015ApJ...800..114S}, but currently lack support from hydrodynamical simulations \citep{2019MNRAS.488.5615S, 2022MNRAS.514.3041Z}.

We caution that Figure \ref{fig:P dist} shows a \textit{very} skewed image due to the overabundance of astrometrically observed systems. Nonetheless, generally the shape of the WD+MS binary period distribution remains debated.
Several studies have suggested that the period distribution of post-mass-transfer systems may be bimodal, with a gap between short-period post-CE systems and long-period non-interacting systems \citep[e.g.,][]{2010ApJS..190..275F,2025arXiv251017957M}.
Yet, Gaia has revealed large populations of systems occupying the apparent  `gap' between the classical short- and long-period regimes \citep{2024MNRAS.529.3729S,2024MNRAS.52711719Y,2024ApJ...970L..11H,2024A&A...686A..61B}. 
Other studies reported log-normal \citep{2011A&A...536A..43N} or log-uniform \citep{2026PASP..138c4202S} distributions for likely PCEBs.

In our catalog, WD+MS binaries span a broad period range, from roughly $0.1$ to $10^{4}$ days. 
Although the distribution shows concentrations at short periods ($P \lesssim 1$ day) and long periods ($P \gtrsim 100$ days), the intermediate regime is also well populated by spectroscopically identified WD+MS binaries.
We emphasize that these distributions are not corrected for observational selection effects and therefore avoid strong quantitative interpretation at this stage.

\subsection{Masses and Mass Ratios \label{ss: mass correlations}}

\begin{figure*}
\centering
\includegraphics[width=\textwidth]{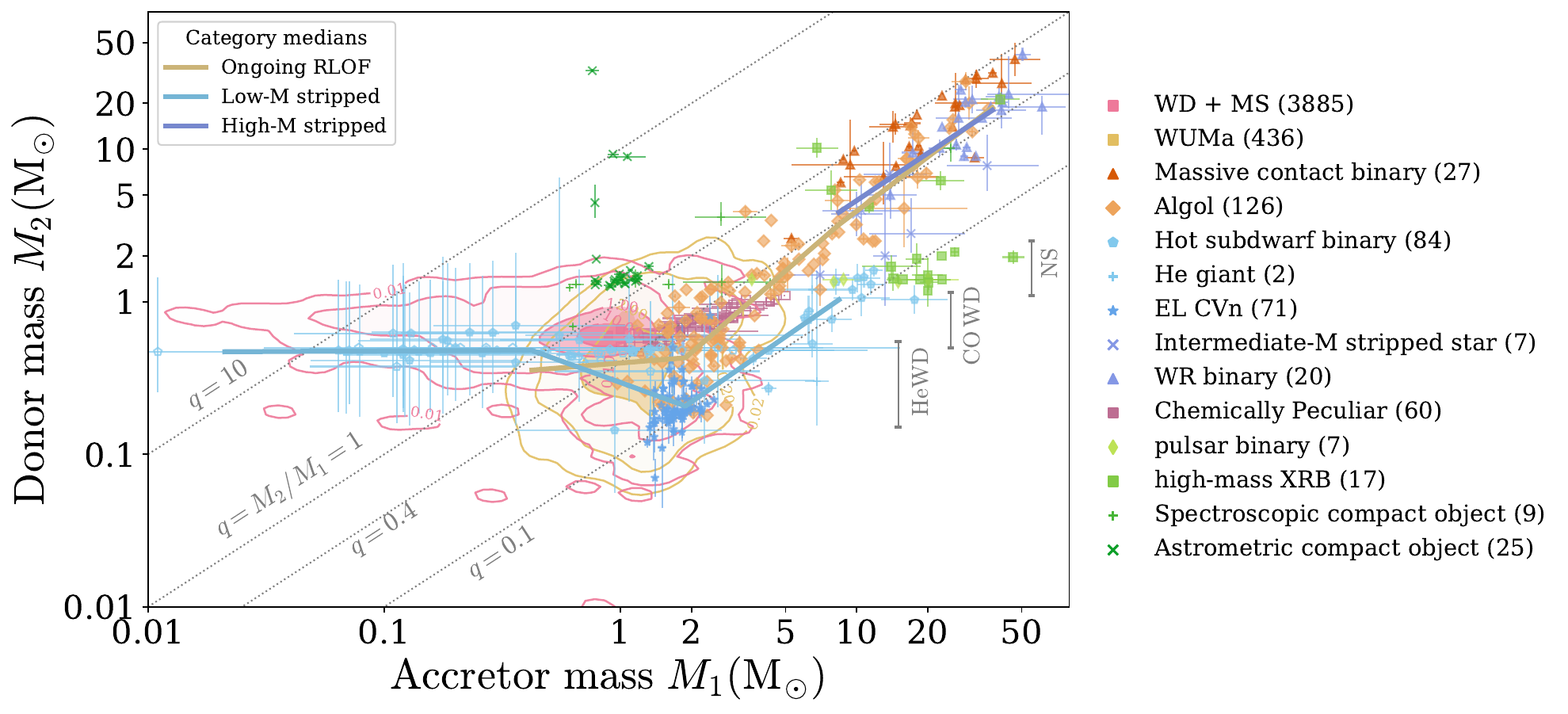}
\caption{The donor ($M_2$) versus accretor ($M_1$) mass for the systems in our catalog, colored by system class. Thick lines show the median $\log_{10}M_2$ in bins of $\log_{10}M_1$ for several overarching categories. WD + MS and W~UMa binaries are shown using a contour KDE for constant bins in $\log M$. Empty symbols indicate inferred or assumed, instead of measured masses.
Grey dotted lines show lines of constant mass ratio. We also indicate approximate masses of typical He WDs ($0.15$-$0.55\Msun$), CO WDs ($0.5$-$1.15\Msun$) and NSs ($1$-$2.5\Msun$).
Post-mass-transfer systems span a wide range of mass ratios within categories. `Ongoing RLOF' systems (Algols and contact binaries) appear to cluster around a fixed mass ratio of $q \simeq 0.4$.
\protect\href{https://github.com/binary-observations/post_mt_catalog/blob/master/code/data_analysis/mass_ratio_Figure.py}{Code used} to generate this plot.
}
\label{fig:M_1_M_2}
\end{figure*}

In Figure~\ref{fig:M_1_M_2} we show the donor mass $M_2$ as a function of the accretor mass $M_1$ for the systems in our catalog, colored by system class. 
We further indicate the median $\log_{10} M_2$ in bins of $\log_{10} M_1$ for each overarching category. Lines of constant mass ratio, $q = M_2/M_1$, are shown with dotted lines for reference. 
Most system classes span a broad range of mass ratios. 
Since the donor is by definition initially more massive than the accretor (since it is the star that evolved first), all post-mass-transfer systems start with $q > 1$ and can only evolve toward lower $q$ via mass transfer; how far below 1 the present-day $q$ lies thus loosly reflects how how (non) conservative the past mass transfer was. 
We will discuss the systems from low to high mass ratios. 

EL CVn systems, together with part of the sdOB and high-mass XRB populations (NS systems with massive companions of $\sim 10$--$20\,\Msun$), lie near $q \simeq 0.1$, consistent with largely conservative mass transfer.
We caution that for several of these classes the donor masses are assumed canonical values or lower limits (due to unknown inclination), which may bias the inferred $q$ and any conclusions about mass-transfer efficiency.
The sdOBs in this region could either be the progenitors of CO WDs or maybe NS progenitors. 

The `Ongoing RLOF' category (Algols and contact binaries) and `High-M stripped' categories show evidence for a preferred mass ratio, as systems cluster around $q \simeq 0.4$.
This is consistent with the findings for massive contact binaries of \cite{2025A&A...695A.223V}, who showed that standard population synthesis models of massive contact binaries overpredict mass ratios near unity because the accretor is too efficiently rejuvenated. 
By limiting core rejuvenation through a modified convective overshooting prescription, they demonstrated that the predicted mass ratio distribution shifts to lower values, in better agreement with the observed sample. 
This suggests that the mass-ratio distribution of contact binaries encodes information about interior mixing processes in accreting stars.
Algol binaries undergo mass ratio reversal where the Roche-lobe filling star is stripped of its outer hydrogen envelope and becomes the less massive component of the binary. The mass ratios (as defined in this work) agree well with the predictions of detailed binary evolution models undergoing mass transfer on the main sequence \citep{2022A&A...659A..98S}. The distribution of mass ratios depends on mass-transfer efficiency; hence, the mass-ratio distribution of Algols and WR+O star binaries can be used as a tracer to empirically constrain the efficiency of mass transfer in binary evolution \citep[e.g.,][]{2007A&A...467.1181D,2011A&A...528A..16V,2025A&A...695A.117N,2025ApJ...990L..51L}. 

White-dwarf-containing systems (i.e., pink populations) show two distinct groupings in mass space. 
One population is concentrated near donor masses of $M_2 \sim 0.2\,\Msun$ and accretor masses of $M_1 \approx 1$--$2\,\Msun$. 
These donor masses are consistent with He WDs, expected from low-mass red giant donors that lose their envelopes before reaching the tip of the giant branch. 
EL CVn systems also lie in this region, which is expected, since they host pre-He WDs.
A second population spans a broad range in $M_1 \sim 0.1-1 \Msun$, with donor masses around $M_2 \sim 1 \Msun$. 
As such, their mass ratios range from $q \approx 1-10$.
The donor masses of these systems are broadly consistent with CO WDs. Part of the sdOB systems (light-blue pentagons) occupy the same region of parameter space, indicating that they are (the progenitors of) lower-mass CO WDs that formed via the sdB channel. \footnote{CO WDs, however, have higher masses and likely descend from AGB donors, skipping the sdB phase.}

Gaia NS systems, i.e., astrometric compact-object binaries with donor masses of $\sim 1-2\Msun$, (dark green crosses) have mass ratios slightly above unity. 
Values of $q > 1$ point toward non-conservative mass transfer, in which the accretor gained relatively little mass.
Gaia BH systems (dark green crosses) are clear outliers, with extreme mass ratios of roughly 10:1 to 30:1 (BH mass to luminous companion mass).

\subsection{Are the astrometric BHs and NSs outliers?\label{ss: are Gaia BHNS outliers}}

As discussed in \S\ref{ss: Gaia_CO}, the astrometric BH systems discovered by Gaia have raised fundamental questions about their formation.
It has been argued that these systems challenge our understanding of binary mass transfer because their extreme mass ratios make it difficult for the binary to avoid a CE phase at mass transfer, while their orbits are simultaneously too wide to be explained by typical post-CE separations but also too close to be comfortably considered as non-interacting systems \citep{2023MNRAS.518.1057E,2023MNRAS.521.4323E,2024A&A...686L...2G}.
When viewed in the context of the broader post-mass-transfer population compiled here, these systems are less anomalous than they appear at face value.

Generally, the astrometric BH and NS systems occupy a similar region of period-eccentricity space as other post-mass-transfer systems, particularly those with high-mass donors.
Although their eccentricities ($0.12 < e < 0.73$) are slightly higher than the overall median eccentricity at their periods ($\sim 200-4000$~d), (which is dominated by low-mass donor systems)  they are consistent with the median eccentricity of the high-mass donor category. 
See dark green crosses in the top panel of Figure~\ref{fig:P_e_diagrams}, and the bottom panels of the same figure.
All of these systems fall within the $90\%$ confidence interval of the median eccentricity for the high-mass donor category, which at these periods is shaped by Wolf--Rayet binaries, pulsar binaries, and high-mass X-ray binaries (excluding the astrometric systems themselves).
More generally, their periods overlap with the period range inhabited by WD binaries and low-mass stripped stars, many of which are potential post-CE systems. 

In mass ratio, the picture is more mixed. The astrometric NS systems have $q = M_\mathrm{CO}/M_\mathrm{comp} \approx 1$ (Figure~\ref{fig:M_1_M_2}), comparable to many WD binaries hosting CO~WDs, whereas the three Gaia BHs are clear outliers with $q \approx 10-30$.
We emphasize, however, that despite their present-day mass ratios, the progenitors of Gaia NSs must have had much more extreme mass ratios (of order $\sim$1:20), which poses similar challenges for avoiding a CE phase, and raises the same questions about how they ended up in their current orbits.

Critically, the same combination of moderately wide orbits and extreme initial mass ratios that makes the Gaia BHs and NSs hard to explain also applies to a subset of the wider Gaia-detected WD+MS population. Systems in the AU-scale ``period gap'' \citep[$P \sim 100$--$1000$~d;][]{2024MNRAS.529.3729S, 2024MNRAS.52711719Y, 2025PASP..137j4205Y} are similarly too wide to be readily explained by standard post-CE evolution, yet with present-day mass ratios that point to highly non-conservative mass transfer at large initial mass ratios. 
\citet{2025PASP..137j4205Y} explicitly find that no standard CE or stable-MT prescription reproduces the observed AU-scale WD+MS population. 
The Gaia BH and NS systems can therefore be viewed as the high-mass extension of a population-wide problem in our understanding of mass transfer at moderately extreme initial mass ratios, rather than as a uniquely compact-object puzzle. 
This parallel framing has been noted in several recent works \citep[e.g.,][]{2025PASP..137j4205Y, 2026arXiv260412839M}.

We caution that the Gaia BH and NS samples are currently small, and the underlying Gaia selection function preferentially detects specific combinations of orbital period, mass ratio, and inclination \citep{2024OJAp....7E..58E,2024MNRAS.529.3729S}. The intrinsic distributions of dormant compact-object binaries in the Milky Way could therefore differ substantially from what we see in our catalog, and conclusions drawn from these subsamples should be treated as suggestive rather than definitive.

We thus conclude that the astrometric BH and NS systems may not be outliers within the broader population of post-mass-transfer binaries. Instead, they may highlight more general shortcomings in our understanding of mass transfer. In particular, they suggest that the separation between post-common-envelope and post-stable mass-transfer systems is not yet well understood.

\section{Conclusions} \label{sec:conclusions}
We have presented a curated, living catalog of  \TotalSystems \, post-mass-transfer binary systems, assembling observational data across sixteen system classes drawn from a wide range of literature sources. 
We focus on systems that have most likely experienced mass transfer from only one star (from the initially more massive to the initially less massive star), with the goal of obtaining the cleanest possible sample for studying binary mass transfer processes.
The catalog, publicly available at \url{https://binary-observations.github.io/post_mt_catalog/} includes orbital periods, eccentricities, component masses, and mass ratios, and is intended as a community resource to support future theoretical and observational work.
For each system class, we reviewed the physical evidence for a mass-transfer origin, the typical observational signatures, and some of the strengths and limitations of the detection methods employed.
Although the main goal of this work is the catalog assembly, we also present an analysis of the period, eccentricity, and masses to showcase how global trends across this catalog can be used to test and inform binary evolution theory.
Our analysis yields four key results:\\

\noindent \textbf{Post-mass-transfer systems are not circular} (top panel of Figure \ref{fig:P_e_diagrams}).
We find non-zero eccentricities at all orbital periods, across system classes and categories. 
Both the median eccentricity and the width of the eccentricity distribution increase with orbital period.
This is consistent with earlier studies of individual classes, including barium stars, post-AGB binaries, and hot subdwarf binaries, which already hinted at this behaviour (see discussion in \S\ref{ss: P-e diagram} and references therein).
Our work extends those results by demonstrating, with the largest and most diverse catalog assembled so far, that it persists across the post-mass-transfer binary population as a whole.
The only exceptions are contact binaries and Algols, but since these are typically fit under the explicit assumption of circular orbits, they provide no independent constraint on the true eccentricity. Conversely, while measurement noise can inflate small individual eccentricities into spuriously non-zero values \citep[e.g.][]{2013A&A...551A..47L}, the population-level increase we report are well above any plausible noise floor and therefore robust against this bias.
The ubiquity of eccentric orbits across such diverse post-interaction systems strongly suggests that the standard assumption of efficient tidal circularization prior to, or during, RLOF is incomplete. Physical processes currently missing or poorly treated in binary evolution models, including phase-dependent mass transfer, tidally enhanced winds, and interactions with circumbinary disks are likely responsible. 
The period-eccentricity diagram thus emerges as a powerful, unifying diagnostic of post-mass-transfer evolution. \\

\noindent \textbf{The $e(\log P)$ relation depends on the donor's progenitor mass} (bottom panels of Figure \ref{fig:P_e_diagrams}).
Dividing the catalog into presumably low-mass and high-mass donor systems reveals a systematic offset in median eccentricity. 
For low-mass donors, the median eccentricity rises from 
$\sim 0.05$ at $\log P \sim -0.5$ to $\sim 0.2$ at $\log P \sim 3.5$, 
while for high-mass donors it increases from $\sim0$ to $\sim 0.7$ over the same range. 
This difference is largely driven by compact-object binaries, which exhibit systematically higher eccentricities than all other categories. This can potentially be understood as the effect of natal kicks imparted during the supernova that formed the neutron star or black hole. 
We caution that the high-mass donor sample is substantially smaller, making these trends less statistically robust, especially at the low-and high period extremes of the catalog.\\

\noindent \textbf{Period distributions are broad and overlapping across categories} (Figure \ref{fig:P dist}).
Contrary to the simple theoretical expectation of a clean separation between short-period post-CE systems, longer-period post-stable-mass-transfer systems, and non-interacting systems at the widest orbital periods, we find that most evolutionary categories span several decades in $P$ and show significant overlap. 
The only category clearly confined to short periods is ongoing Roche-lobe overflow ($P \lesssim 10$days). 
This overlap may reflect genuine degeneracy between the outcomes of CE and stable mass-transfer evolution, or it may result from observational selection effects that blur intrinsic features in the period distribution. 
Disentangling these plausible causes will require completeness-corrected samples.
We stress that this caveat applies to the post-mass-transfer population as a whole: within homogeneous, well-selected system classes such as sdB+MS (\S\ref{ss: SdOB}) and WD+MS binaries (\S\ref{ss: WD + MS}), the period distribution can still be a powerful diagnostic of mass-transfer history. \\

\noindent \textbf{Post-mass-transfer systems span a wide range of mass ratios} (Figure \ref{fig:M_1_M_2}).
With the exception of ongoing Roche-lobe overflow systems (Algols and contact binaries, which cluster around $q \approx 0.4$), all categories show broad distributions in the donor-accretor mass plane. This diversity reflects the wide range of initial conditions and mass-transfer efficiencies operating across these populations. \\

\noindent \textbf{The astrometric Gaia BH and NS systems are not clear outliers with respect to the broader post-mass-transfer population} (Figures~\ref{fig:P_e_diagrams}, \ref{fig:M_1_M_2}). 
In period-eccentricity space they overlap with other interacting and post-interaction binaries, particularly high-mass donor systems, and their eccentricities fall within the observed range of this population. 
While the Gaia BH systems are extreme in present-day mass ratio, the NS systems are comparable to CO-WD binaries. Their progenitors would have required similarly extreme initial mass ratios, implying comparable challenges for post-CE evolution. 
We argue that the Gaia BH and NS systems share these challenges with a subset of the Gaia-detected WD+MS population in the AU-scale ``period gap'' (\S\ref{ss: WD + MS}), which is similarly difficult to reconcile with standard CE or stable-MT predictions at large initial mass ratios. The puzzle posed by the Gaia BHs and NSs may therefore be a high-mass extension of a population-wide problem rather than a uniquely compact-object phenomenon.
We caution, however, that the Gaia BH and NS samples are small and subject to selection effects that preferentially detect specific orbital configurations.
Overall, they appear to be part of a broader picture highlighting limitations in our treatment of binary mass transfer, rather than as isolated outliers.\\

Taken together, these results highlight persistent tensions between current binary evolution theory and observations, and underscore the value of a unified, multi-population approach. 
Future work will aim to apply completeness corrections to the period and mass-ratio distributions, expand the catalog with newly discovered systems and use the catalog as a benchmark for population synthesis models. 
Ultimately, this work is intended as a first foundation for bridging the gap between theoretical predictions and the growing body of observational data on interacting binary systems.


\begin{acknowledgments}
This paper originated at the Center for Computational Astrophysics `Stable Mass Transfer workshop' held in May 2025. We thank the Simons Foundation for providing support and resources for the meeting.
LvS is supported by VI.Veni.242.115, Grant ID \url{https://doi.org/10.61686/XVIAV86753}. TS acknowledges support from the European Research Council (ERC) under
the European Union's Horizon 2020 research and innovation programme
(grant agreement No.~101164755, METAL) and the Israel Science Foundation
(ISF, grant No.~0603225041). NY acknowledges support from the Ezoe Memorial Recruit Foundation scholarship. A.L. acknowledges support from the NSERC and is funded through a NSERC Canada Graduate Scholarship—Doctoral. A.L. is also supported by the Data Sciences Institute
at the University of Toronto through grant number DSIDSFY3R1P02.
\end{acknowledgments}


\section*{Data Availability}
The catalog presented in this work available at Zenodo \citep[10.5281/zenodo.20441987][]{vanson_2026_postmt_zenodo}, as well as a live version with interactive figures and data downloads at \url{https://binary-observations.github.io/post_mt_catalog/}.
The code used to generate the catalog and figures in this paper are available at \url{https://github.com/binary-observations/post_mt_catalog}.

\software{
We used the following software packages: \texttt{astropy} \citep{2013A&A...558A..33A,2018AJ....156..123A,2022ApJ...935..167A}, \texttt{Jupyter} \citep{2007CSE.....9c..21P}, 
\texttt{matplotlib} \citep{Hunter:2007}, 
\texttt{numpy} \citep{numpy},
and  \texttt{python} \citep{python}.
This research has made use of the Astrophysics Data System, funded by NASA under Cooperative Agreement 80NSSC21M00561.
Software citation information aggregated using \texttt{\href{https://www.tomwagg.com/software-citation-station/}{The Software Citation Station}} \citep{software-citation-station-paper,software-citation-station-zenodo}.
This research has made use of the SIMBAD database \citep{2000A&AS..143....9W}, and the VizieR catalogue access tool \citep[DOI : 10.26093/cds/vizier][]{2000A&AS..143...23O} operated at CDS, Strasbourg, France. 
}

\begin{contribution}
    LvS led the general catalog assembly (including code development and data curation), and the bulk of the manuscript (abstract, introduction, catalog construction, correlation analysis, and conclusions). 
    The compilation of the catalog and the writing of subsections for individual binary classes were carried out collaboratively by the co-authors, who each contributed substantially to these components:
    NY: astrometric and spectroscopic WD+MS binaries, barium stars and post-AGB binaries;
    PN: astrometric and spectroscopic BH/NS+MS binaries, young pulsar binaries, and X-ray binaries;
    TS: Wolf Rayet binaries and He giants;
    KS: Algols and contact binaries;
    AL: stripped stars and He giants;
    EL: Blue straggler binaries;
    HS: Contact binaries;
    OP: detailed review of the catalog and manuscript, particularly the hot subdwarf, chemically peculiar, post-AGB, and WD+MS subsections
    LvS: hot subdwarf binaries, post-AGB binaries, chemically peculiar binaries, and EL CVn systems.
    All authors furthermore contributed to the concept design of the catalog and the writing and editing of the manuscript, and provided feedback on the analysis and interpretation of the results.
\end{contribution}


\appendix

\section{Other post-mass transfer objects, not included in this work \label{sec: not incl}}

As discussed in \S \ref{sec:catalog_construction}, our catalog focuses on systems that have experienced only a single phase of mass transfer, i.e. systems in which only the initially more massive star has transferred a significant amount of mass. 
This allows us to isolate the imprint of mass-transfer with the least ambiguity. 
We therefore exclude systems that have likely undergone multiple interaction phases. However, such systems remain important for understanding the full picture of binary evolution, and we briefly highlight them here.
Specifically, the following classes are likely to have undergone an additional mass transfer phase from the initially less massive star:

\begin{itemize}[nolistsep]
\item Double compact object binaries: Including double white dwarf (WD+WD), double neutron star (NS+NS) binaries, and WD + NS binaries.
Both stars must have evolved and lost their envelopes, implying a mass-transfer episode from each component. 

 \item Cataclysmic variables (CVs; WD + MS): The WD accretes via Roche-lobe overflow from a low-mass companion, indicating a second mass-transfer phase after the WD has formed.
 
 \item Low-mass X-ray binaries (LMXBs): A NS or BH accretes from a low-mass donor. 
 LXMBs are currently experiencing stable mass transfer, and have likely already undergone an episode of common-envelope evolution in the past \citep[][]{2006csxs.book..623T}, and thus represent ongoing second-phase mass transfer.

\item Symbiotic binaries: a compact object (NS or WD) accretes from the wind of an evolved red-giant or AGB companion \citep[e.g.,][]{2019MNRAS.485..851Y}. In the low-mass case the accretor is a WD, representing a later interaction phase following the formation of the WD. The high-mass symbiotic X-ray binaries (SyXBs) host a NS accreting from a red giant \citep[e.g.,][]{2006ApJ...641..479H,2019ApJ...872...43H,2024PASP..136g4202N}. It is unclear whether these systems are still pre-`second' mass transfer. Since they are edge-cases we exclude them from the current version of our catalog.

\end{itemize}

We also exclude photometric compact-object binaries (ellipsoidal variables).  
These should show coherent, double-peaked modulation from tidal distortion, indicating close, interaction-shaped binaries, typically found in photometric surveys and confirmed spectroscopically (period, RV amplitude).  
No BH or NS has been discovered via this method to date 
\citep[e.g.,][]{2023A&A...674A..19G, 2023MNRAS.524.4367N, 2023AcA....73..197K, 2024AJ....168...44O, 2024OJAp....7E..24R, 2025A&A...695A.210G}. 
This is because such binaries are rare, and easily outnumbered by contact binaries, whose light curves can mimic ellipsoidal variability and serve as a major contaminant.


\bibliographystyle{aasjournalv7}
\bibliography{main,software_citations}



\end{document}